\documentclass[twocolumn]{aastex63}

\turnoffeditone

\received{July 31, 2020}
\revised{September 3, 2020}
\accepted{September 16, 2020}

\submitjournal{\apj}

\shorttitle{Probing the Nature of High Redshift Weak Emission Line Quasars}
\shortauthors{Andika et al.}
\begin{document}

\title{Probing the Nature of High Redshift Weak Emission Line Quasars:\\
A Young Quasar with a Starburst Host Galaxy}

\correspondingauthor{Irham Taufik Andika}
\email{andika@mpia.de}

\author[0000-0001-6102-9526]{Irham Taufik Andika}
\affiliation{Max-Planck-Institut f\"{u}r Astronomie, K\"{o}nigstuhl 17, D-69117 Heidelberg, Germany}
\affiliation{International Max Planck Research School for Astronomy \& Cosmic Physics at the University of Heidelberg}

\author[0000-0003-3804-2137]{Knud Jahnke}
\affiliation{Max-Planck-Institut f\"{u}r Astronomie, K\"{o}nigstuhl 17, D-69117 Heidelberg, Germany}

\author[0000-0003-2984-6803]{Masafusa Onoue}
\affiliation{Max-Planck-Institut f\"{u}r Astronomie, K\"{o}nigstuhl 17, D-69117 Heidelberg, Germany}

\author[0000-0002-2931-7824]{Eduardo Ba\~{n}ados}
\affiliation{Max-Planck-Institut f\"{u}r Astronomie, K\"{o}nigstuhl 17, D-69117 Heidelberg, Germany}

\author[0000-0002-5941-5214]{Chiara Mazzucchelli}
\affiliation{European Southern Observatory, Alonso de C\'ordova 3107, Vitacura, Regi\'on Metropolitana, Chile}

\author[0000-0001-8695-825X]{Mladen Novak}
\affiliation{Max-Planck-Institut f\"{u}r Astronomie, K\"{o}nigstuhl 17, D-69117 Heidelberg, Germany}

\author[0000-0003-2895-6218]{Anna-Christina Eilers}\altaffiliation{NASA Hubble Fellow}
\affiliation{MIT Kavli Institute for Astrophysics and Space Research, 77 Massachusetts Ave., Cambridge, MA 02139, USA}
\affiliation{Max-Planck-Institut f\"{u}r Astronomie, K\"{o}nigstuhl 17, D-69117 Heidelberg, Germany}

\author[0000-0001-9024-8322]{Bram P. Venemans}
\affiliation{Max-Planck-Institut f\"{u}r Astronomie, K\"{o}nigstuhl 17, D-69117 Heidelberg, Germany}

\author[0000-0002-4544-8242]{Jan-Torge Schindler}
\affiliation{Max-Planck-Institut f\"{u}r Astronomie, K\"{o}nigstuhl 17, D-69117 Heidelberg, Germany}

\author[0000-0003-4793-7880]{Fabian Walter}
\affiliation{Max-Planck-Institut f\"{u}r Astronomie, K\"{o}nigstuhl 17, D-69117 Heidelberg, Germany}

\author[0000-0002-9838-8191]{Marcel Neeleman}
\affiliation{Max-Planck-Institut f\"{u}r Astronomie, K\"{o}nigstuhl 17, D-69117 Heidelberg, Germany}

\author[0000-0003-3769-9559]{Robert A. Simcoe}
\affiliation{MIT Kavli Institute for Astrophysics and Space Research, 77 Massachusetts Ave., Cambridge, MA 02139, USA}

\author[0000-0002-2662-8803]{Roberto Decarli}
\affiliation{INAF--Osservatorio di Astrofisica e Scienza dello Spazio di Bologna, Via Gobetti 93/3, I-40129, Bologna, Italy}

\author[0000-0002-6822-2254]{Emanuele Paolo Farina }
\affiliation{Max-Planck-Institut f\"{u}r Astrophysik, Karl-Schwarzschild-Stra{\ss}e 1, D-85748, Garching bei M\"{u}nchen, Germany}

\author[0000-0003-1733-9281]{Victor Marian}
\affiliation{Max-Planck-Institut f\"{u}r Astronomie, K\"{o}nigstuhl 17, D-69117 Heidelberg, Germany}
\affiliation{International Max Planck Research School for Astronomy \& Cosmic Physics at the University of Heidelberg}

\author[0000-0001-9815-4953]{Antonio Pensabene}
\affiliation{INAF--Osservatorio di Astrofisica e Scienza dello Spazio di Bologna, Via Gobetti 93/3, I-40129, Bologna, Italy}
\affiliation{Dipartimento di Fisica e Astronomia, Alma Mater Studiorum, Universit\`a di Bologna, Via Gobetti 93/2, I-40129 Bologna, Italy}

\author[0000-0003-4063-5126]{Thomas M. Cooper}
\affiliation{The Observatories of the Carnegie Institution for Science, 813 Santa Barbara Street, Pasadena, CA 91101, USA}

\author[0000-0003-0006-8681]{Alejandra F. Rojas}
\affiliation{Centro de Astronomía (CITEVA), Universidad de Antofagasta, Avenida Angamos 601, Antofagasta, Chile}



\begin{abstract}
We present the discovery of PSO J083.8371+11.8482, a weak emission line quasar with extreme star formation rate at $z=6.3401$.
This quasar was selected from Pan-STARRS1, UHS, and unWISE photometric data.
Gemini/GNIRS spectroscopy follow-up indicates a Mg\,\textsc{ii}-based black hole mass of $M_\mathrm{BH}=\left(2.0^{+0.7}_{-0.4}\right)\times10^9$~$M_\odot$ and an Eddington ratio of $L_\mathrm{bol}/L_\mathrm{Edd}=0.5^{+0.1}_{-0.2}$, in line with actively accreting supermassive black hole (SMBH) at $z\gtrsim6$.
HST imaging sets strong constraint on lens-boosting, showing no relevant effect on the apparent emission. The quasar is also observed as a pure point-source with no additional emission component.
The broad line region (BLR) emission is intrinsically weak and not likely caused by an intervening absorber. We found rest-frame equivalent widths of EW$(\rm Ly\alpha+N\,\textsc{v})_{rest}=5.7\pm0.7$~\AA, EW$(\rm C\,\textsc{iv})_{rest}\leq5.8$~\AA~(3-sigma upper limit), and EW$(\rm Mg\,\textsc{ii})_{rest}=8.7\pm0.7$~\AA.
A small proximity zone size ($R_\mathrm{p}=1.2\pm0.4$~pMpc) indicates a lifetime of only $t_\mathrm{Q}=10^{3.4\pm0.7}$ years from the last quasar phase ignition.
ALMA shows extended [C\,\textsc{ii}] emission with a mild velocity gradient. The inferred far-infrared luminosity ($L_\mathrm{FIR}=(1.2\pm0.1)\times10^{13}\,L_\odot$) is one of the highest among all known quasar hosts at $z\gtrsim6$.
Dust and [C\,\textsc{ii}] emissions put a constraint on the star formation rate of SFR~$=900$--$4900~M_\odot\,\mathrm{yr^{-1}}$, similar to that of hyper-luminous infrared galaxy.
Considering the observed quasar lifetime and BLR formation timescale, the weak-line profile in the quasar spectrum is most likely caused by a BLR which is not yet fully formed rather than continuum boosting by gravitational lensing or a soft continuum due to super-Eddington accretion.

\end{abstract}



\section{Introduction} 
\label{sec:intro}

Quasars\footnote{We use the terms ``quasar'' and ``QSO'' interchangeably throughout this paper.} are the most luminous non-transient sources in the Universe and they are very rare with number density of only $\lesssim 1$~Gpc$^{-3}$ \citep{2019arXiv191105791I}.
Their spectra provide important insights into the properties of the intergalactic medium (IGM) at the later stages of the epoch of reionization (EoR) which is thought to be completed by $z\sim5.5$ (e.g.\ \citealt{2006AJ....132..117F,2015MNRAS.447.3402B,2016ASSL..423..187M}).
They are also excellent probes for understanding the build-up of the first supermassive black holes (SMHBs; e.g.\ \citealt{2010A&ARv..18..279V}) as well as the metal enrichment of the first galaxies and of their surroundings (e.g.\ \citealt{2017ApJ...850..188C,2017MNRAS.470.1919B,2019A&ARv..27....3M,2020arXiv200616268O}).
Identification and detailed characterization of more quasars at this epoch will provide us with better constraints on the physical properties of the Universe at end of the transition phase from neutral to ionized hydrogen.

The existence of billion-solar-mass SMBHs at $z\gtrsim6$ (e.g.\ \citealt{2011Natur.474..616M,2015Natur.518..512W,2018Natur.553..473B,2017ApJ...849...91M,2020ApJ...897L..14Y}) puts stringent constraints on the formation and early growth models of the first SMBHs and host galaxies within the first Gyr after the Big Bang (see \citealt{2019arXiv191105791I} for a recent review).
This challenges standard SMBH formation models, which start from a stellar seed mass black hole that grows via Eddington-limited accretion (e.g.\ \citealt{2010A&ARv..18..279V,2012Sci...337..544V}). 
Among the current theoretical scenarios accounting for the formation of the observed SMBHs are: the growth from massive seed black holes ($\gtrsim10^4~M_\odot$) through direct collapse channel (e.g.\ \citealt{2006MNRAS.370..289B,2014MNRAS.443.2410F,2016MNRAS.463..529H,2017MNRAS.471.4878S,2019MNRAS.486.2336D}), lower-mass seeds ($\lesssim10^{2-3}~M_\odot$) with Eddington limited or even super-Eddington accretion and very rapid growth (e.g.\ \citealt{2005ApJ...628..368O,2009ApJ...696.1798T,2016MNRAS.459.3738I}), or presence of radiatively inefficient accretion modes (e.g.\ \citealt{2017ApJ...836L...1T,2019ApJ...884L..19D}).
Recent identification of young quasars with estimated lifetimes of only $t_\mathrm{Q} < 10^{4-5}$ years at $z\sim6$ \citep{2017ApJ...840...24E,2018ApJ...867...30E,2020MNRAS.493.1330D,2020arXiv200201811E} imposed additional tensions with respect to the standard growth models of SMBHs.
These young quasars are identified based on their small proximity zones, which is the region of enhanced Ly$\alpha$ forest transmission produced by ionizing radiation from the central quasar and before the onset of the \cite{1965ApJ...142.1633G} absorption trough (e.g.\ \citealt{2017ApJ...840...24E}).

At lower-redshift ($z\sim3-5$), a notable group of quasars  shows exceptionally weak ultraviolet broad emission lines, originally discovered in the Sloan Digital Sky Survey spectra and systematically investigated by \cite{2009ApJ...699..782D}.
These so called weak emission line quasars (WLQs) are defined as having rest-frame equivalent width of 
EW$(\rm Ly\alpha~\lambda1216+N\,\textsc{v}~\lambda1240)_{rest} < 15.4$\,\AA~and/or EW$(\rm C\,\textsc{iv}~\lambda1549)_{rest} < 10$\,\AA. Meanwhile, the EW($\rm C\,\textsc{iv}$) of normal quasars follow a log-normal distribution with a mean of $\langle\mathrm{EW(C\,\textsc{iv})}\rangle = 42^{+25}_{-16}$\,\AA\ \citep{2009ApJ...699..782D}. Hence, WLQs are the $3\sigma$ outliers at the low-end of this distribution.
Several theories has been proposed to explain the WLQ phenomenon and according \cite{2015ApJ...805..123P}, they fall into two broad categories: (1) soft ionizing continuum idea and (2) anemic broad emission line region (BLR) model.
In the soft ionizing continuum idea, one might expect that the BLR is less photoionized so the produced broad emission lines are weak, probably because of: (i) inefficient photoionizing photons due to an extremely high accretion rate \citep{2007ApJS..173....1L,2007ApJ...663..103L}, (ii) low accretion rate in the very massive black hole which leads to radiatively inefficient cold accretion disk \citep{2011MNRAS.417..681L}, or (iii) accretion disk produced high-energy photons are absorbed by shielding materials \citep{2011ApJ...736...28W,2015ApJ...805..122L,2018MNRAS.480.5184N}.
On the other hand, anemic BLR model suggests that the BLR itself could be unusually gas deficient \citep{2010ApJ...722L.152S,2012MNRAS.420.2518N}, possibly if the quasar is in an early phase of accretion and the BLR has not yet fully formed \citep{2010MNRAS.404.2028H,2014A&A...568A.114M}.
It is then critical to study the intrinsic properties of WLQs and environment of their host galaxies at the highest accessible redshifts to test the evolutionary scenario.

Up to now, there have been around 270 quasars discovered at $z>6$, mostly found thanks to large-area or deep sky surveys (e.g.\ \citealt{2006AJ....132..117F,2010AJ....139..906W,2011Natur.474..616M,2013ApJ...779...24V,2015ApJ...801L..11V,2016ApJ...833..222J,2016ApJS..227...11B,2016ApJ...828...26M,2017ApJ...849...91M,2017MNRAS.468.4702R,2017ApJ...839...27W,2018PASJ...70S..35M,2018ApJS..237....5M,2019MNRAS.487.1874R,2019AJ....157..236Y,2019ApJ...884...30W,2019MNRAS.484.5142P}).
Among those discovered $z>6$ quasars, there are only $\sim20$ of them which are identified as WLQs \citep{2016ApJS..227...11B,2019ApJ...873...35S}.
To increase sampling points of reionization at $z>6$, and to investigate the relationship of rare WLQs and cases of very young age quasars which allows us to handle on early modes of growth, an increase in the quasar sample size at this early epoch is critical.

In this paper, we report the discovery of PSO\,J083.8371+11.8482 (hereafter PSO J083+11) at $z = 6.34$, as part of our effort to expand the number of known quasars at $z>6$. 
In order to investigate the physical properties of this quasar, its host galaxy and its environment, we performed an extensive, multi-wavelength (from optical/near-infrared to sub-millimeter) campaign with state of the art facilities, that we present here.
The primary data used for initial candidates selection are outlined in Section~\ref{sec:phot_catalog}, while our method for selecting quasars via spectral energy distribution modeling is described in Section~\ref{sec:sed_fit}. Then, we report the spectroscopic follow-up data in Section~\ref{sec:spec_obs} for confirming the quasar nature of PSO J083+11 and derive black hole properties. High resolution Hubble Space Telescope (HST) near-infrared imaging to test whether gravitational lensing affects the quasar's apparent emission is presented in Section~\ref{sec:hst}. After that, in Section \ref{sec:weak_line}, proximity zone size and quasar lifetime calculations are presented. The host galaxy properties from sub-millimeter observation based on Atacama Large Millimeter/sub-millimeter Array (ALMA) are explored in Section~\ref{sec:alma}. Section~\ref{sec:discussion} discusses possible physical processes that drive the weakness of PSO J083+11 BLR emission. We close by summarizing the paper and its conclusions in Section~\ref{sec:summary}.

We use AB zeropoints for all magnitudes written in this paper. We further assume a flat $\Lambda$CDM cosmology with $\Omega_\mathrm{m}=0.3$, $\Omega_\Lambda=0.7$, and $H_0 = 70~\rm km~s^{-1}~Mpc^{-1}$ for all of physical measurements.
Using this assumption, at $z = 6.3401$ the age of Universe is 0.852 Gyr and an angular scale of $\theta = 1\arcsec$ corresponds to a proper transverse separation of 5.6 kpc.

\section{Initial Candidate Selection from Public Multiband Data}
\label{sec:phot_catalog}

Our quasar candidate selection is a two part process: 1) pre-selection of candidates from multiband photometric data, 2) modeling of the spectral energy distribution to derive relative probabilities for a candidate to be a quasar or contaminant.

We started the first part of our quasar search by exploiting the Pan-STARRS1 survey, following and expanding the selection by \cite{2017ApJ...849...91M} to focus on the redshift range $6.3 \leq z \leq 7.1$. Then, we added various infrared photometric data points from public surveys to help us classify and estimate the photometric redshift of each candidate. The catalog photometry was corrected for Galactic reddening by using \texttt{Bayestar19} dustmap \citep{2019ApJ...887...93G} and the \cite{1999PASP..111...63F} reddening law. 
\edit1{Then, we cross-matched with a list of late M stars along with L and T dwarfs (hereafter MLT dwarfs) from \cite{2018ApJS..234....1B} and quasars from \cite{2019arXiv191205614F} to exclude those known objects from our list of $z>6$ quasar candidates.}
Finally, we carried out follow-up spectroscopy of promising targets as the final confirmation of their nature.
We will provide the technical details for each step in the following section.

\subsection{Main optical catalog} 
\label{sec:ps1_catalog}

We use the Pan-STARRS1 catalog internal release version (PS1 version 3.4; \citealt{2016arXiv161205560C}) as the main data for the initial quasar candidate selection. The 5$\sigma$ limiting magnitudes of this stack catalog are $g=23.3$, $r=23.2$, $i=23.1$, $z=22.3$, $y=21.3$. 
Due to intervening intergalactic medium (IGM), the $z\gtrsim6.2$ quasar's flux on the blue side of Ly$\alpha$ is expected to be heavily absorbed, creating a strong break in flux. This will also make them basically undetected, i.e.\ have very low signal-to-noise ratios (S/N), in all bands bluer than $z$-band at the PS1 limiting magnitude.
For the selection, we require candidates to be detected in the PS1 $y$-band and have very red colors. In summary, the criteria used are:
\begin{eqnarray}
    \label{eq:low_snr} \mathrm{S/N}(g_\mathrm{PS1},~r_\mathrm{PS1},~i_\mathrm{PS1}) < 8 \\
    \mathrm{S/N}(y_\mathrm{PS1}) > 7 \\
    \mathrm{S/N}(z_\mathrm{PS1}) \geq 3~\mathrm{and}~z_\mathrm{PS1} - y_\mathrm{PS1} > 1.2 \\
    \nonumber \mathrm{or} \\
    \nonumber \mathrm{S/N}(z_\mathrm{PS1}) < 3~\mathrm{and}~z_\mathrm{PS1,lim} - y_\mathrm{PS1} > 1.2 \\
    \edit1{y_\mathrm{PS1} > 15}     \label{eq:y_maglim}
\end{eqnarray}
where all measured fluxes and magnitudes are based on point spread function (PSF) photometry, unless stated otherwise. \edit1{The magnitude limit criterion in Equation \ref{eq:y_maglim} is used to exclude unusually bright objects or spurious sources.}

However, quasars of extreme brightness could still be detected in some PS1 ``dropout'' bands. This might be because of their intrinsically high luminosity (e.g.\ \citealt{2015Natur.518..512W}) or their apparent fluxes are boosted by gravitational lensing events (e.g.\ \citealt{2019ApJ...870L..11F,2020ApJ...891...64F}). Moreover, most strongly lensed quasars would be removed as candidates due to Equation~\ref{eq:low_snr} because the massive lensing galaxy at intermediate redshift would contribute flux in those bands. Hence, additional criterion is applied if Equation~\ref{eq:low_snr} is not fulfilled to actually find this population:
\begin{equation}
    (g_\mathrm{PS1},~r_\mathrm{PS1},~i_\mathrm{PS1}) - y_\mathrm{PS1} > 3.0
\end{equation}

There are some useful parameters in the PS1 catalog that can be utilized to remove most of the contaminants. The first one is to exclude objects showing extended morphology and only choose point-like or very compact sources by requiring
\begin{eqnarray}
    |y_\mathrm{PS1,aper} - y_\mathrm{PS1}| < 0.5,
\end{eqnarray}
where $y_{\mathrm {PS1,aper}}$ and $y_{\mathrm {PS1}}$ are PS1 catalog aperture and PSF magnitudes of stacked images, respectively.
This cutoff threshold was chosen based on tests of spectroscopically confirmed galaxies and stars, following \cite{2016ApJS..227...11B}. 
Second, the measured PSF magnitudes also need to be consistent with each other:
\begin{eqnarray}
    |y_\mathrm{PS1,stk} - y_\mathrm{PS1,wrp}| < 0.5,
\end{eqnarray}
where ``stk'' and ``wrp'' denote photometry from stacked images and the mean photometry of single-epoch images, respectively.
Lastly, the expected weighted PSF flux is required to be located in good pixels to a percentage of 85\% or more,  i.e., $\rm PSF\_QF>0.85$ in the catalog. By using aforementioned criteria, we obtained $\sim17$~millions candidates at the preliminary selection step (see Table \ref{tab:selection}).

Some areas of PS1 are also covered by the Dark Energy Spectroscopic Instrument Legacy Imaging Surveys Data Release 8 (DELS; \citealt{2019AJ....157..168D}) and the Dark Energy Survey Data Release 1 (DES; \citealt{2018ApJS..239...18A}). The advantages of DELS and DES are that they reach magnitude levels $\sim1$\,mag fainter compared to PS1.
DELS\footnote{\url{http://legacysurvey.org/}} is conducted using imaging data from three different telescopes, covering $\sim 14\,000$ square degrees of extragalactic sky visible from the northern hemisphere ($-18\deg < \mathrm{Dec} < +84\deg$) in three optical bands. These data reach 5-sigma depths of $g=24.0$, $r=23.4$ and $z=22.5$ mag \citep{2019AJ....157..168D}.
On the southern hemisphere, DES\footnote{\url{https://www.darkenergysurvey.org/}} is utilizing the Dark Energy Camera mounted on the Cerro Tololo Interamerican Observatory 4--m Blanco telescope to image $\sim 5000$ square degrees of southern Galactic cap region in five broad bands. The median coadded catalog depth for a 1\farcs95 diameter aperture at S/N = 10 is $g=24.33$, $r=24.08$, $i=23.44$, $z=22.69$, and $Y= 21.44$ mag \citep{2018ApJS..239...18A}.
Hence, the DELS and DES catalogs are cross-matched to our main PS1 catalog with 2\arcsec~radius and we will use their photometric data if available.

Selecting high redshift (``high-$z$'', $z\gtrsim6$) quasar candidates through color criteria becomes complicated due to contamination from other populations: late M stars along with L and T dwarfs (MLT dwarfs); and elliptical galaxies at $z=1$--2 (hereafter ellipticals). These populations have a higher surface density than the target high-$z$ quasars themselves but have similar near-infrared colors. Hence, to reduce the number of MLT dwarfs contaminating the candidate sample, we exclude the sky region around M31 ($7\deg<\rm RA < 14\deg;~37\deg < Dec < 43\deg$) and the Milky Way plane ($|b|<20\deg$). However, with the advantage of \texttt{Bayestar19} \citep{2019ApJ...887...93G} dust maps, we include some candidates in the Galactic plane region if they have reddening of $E(B-V) < 1$ in the selection, as a difference to previous studies that rigorously excluded all regions with $E(B-V) > 0.3$ (e.g.\ \citealt{2016ApJS..227...11B,2017ApJ...849...91M}).

To make a more refined and more robust selection, we need to take advantage of infrared photometry, because it will give us a parameter dimension to distinguish quasars from MLT dwarfs and ellipticals.

\begin{deluxetable}{clrc}

	\tablenum{1}
	\tablecaption{Summary of quasar selections. Each selection step shows the number of candidates and recovered known quasars at $z>6$.}
	\label{tab:selection}
	\tablehead{
		\colhead{Step} & \colhead{Selection} & \colhead{Candidates} & \colhead{Known} \\
		& & & \colhead{Quasars}
	}	

	\startdata
	1 & Initial query with: & & \\
	  & $\rm PSF\_QF>0.85$ & & \\
      & S/N$(y_{\rm PS1}) \geq 5$ & & \\
      & $y_{\rm PS1} > 15$ & & \\
	  & $z_{\rm PS1} - y_{\rm PS1} > 1.0$ & & \\
	  & Excluding Galactic plane &  &  \\
	  & Excluding M31 & 17170000 & 27 \\
	2 & Detection in NIR or MIR & 5631392 & 27 \\
	3 & S/N$(y_{\rm PS1}) \geq 7$ & 3627100 & 27 \\	
    4 & $|y_\mathrm{PS1,aper} - y_\mathrm{PS1}| < 0.5$ & 1158750 & 24 \\
    5  & $|y_\mathrm{PS1,stk} - y_\mathrm{PS1,wrp}| < 0.5$ & 1059373 & 24\\    
	6 & $z_{\rm PS1} - y_{\rm PS1} > 1.2$ & 367190 & 22\\
    7 & S/N$(g_\mathrm{PS1},~r_\mathrm{PS1},~i_\mathrm{PS1}) < 8$, or &  & \\
      & $(g_\mathrm{PS1},~r_\mathrm{PS1},~i_\mathrm{PS1}) - y_\mathrm{PS1} > 3$ & 74374 & 22\\
	8 & SED fitting & 7808 & 20 \\
	9 & Forced photometry & 1263 & 20 \\
	10 & Visual inspection & 155 & 20\\
	\enddata
	\tablecomments{Our method is most sensitive to select quasars at $6.3\leq z \leq 7.1$ due to step 1 requirements.}
\end{deluxetable}

\subsection{Public infrared data} \label{sec:ir_catalog}
We take advantage of public surveys when their sky footprint overlapped with that of PS1. In this case, near-infrared (NIR) photometry for the northern hemisphere was taken from the UKIRT Infrared Deep Sky Survey DR10 (UKIDSS; \citealt{2007MNRAS.379.1599L}) and the UKIRT Hemisphere Survey DR1 (UHS; \citealt{2018MNRAS.473.5113D}),  while in the southern hemisphere we used the Vista Hemisphere Survey DR6 (VHS; \citealt{2013Msngr.154...35M}) with 2\arcsec~cross-matching radius. The $J$- as well as $H$- and $K$-band magnitudes are very useful to discriminate between quasars and MLT dwarfs \citep{2017ApJ...849...91M,2019AJ....157..236Y,2019ApJ...884...30W}. This is also an efficient method to remove spurious sources like cosmic rays that are usually only detected in one survey and not in the other.

In addition, mid-infrared (MIR) data were taken from the unWISE catalog \citep{2019ApJS..240...30S} which contains roughly two billion objects, observed by the Wide-field Infrared Survey Explorer (WISE; \citealt{2010AJ....140.1868W}) over the whole sky. 
The $W1$~(3.4~$\mu$m) and $W2$~(4.6~$\mu$m) of WISE photometric bands are useful to separate quasars from MLT dwarfs (e.g., the $W1-W2$ color).
The advantage of unWISE compared to the original WISE catalog (AllWISE) is significantly deeper imaging data and improved  source extraction in crowded regions \citep{2019ApJS..240...30S}. A 3\arcsec~cross-match radius to our main catalog was applied.
\edit1{By default, we defined that a source is detected in the NIR/MIR if the catalog entry of the corresponding flux measurements is not empty, i.e. there is a match within the cross-matching radius.}

\section{Quasar Search via Spectral Energy Distribution Modeling} \label{sec:sed_fit}

The second part of our selection method is our own implementation of fitting the full available spectral energy distribution (SED) to templates of quasars and the main contaminants to fully exploit the multi-wavelength data. The end result of this procedure is to estimate both the photometric redshift and the probability of each source being a quasar or contaminant.

We used all MLT dwarf spectra observed by \cite{2014ASInC..11....7B} which are stored in the SpeX Prism Library\footnote{\url{http://pono.ucsd.edu/~adam/browndwarfs/spexprism/library.html}}. This contains $\sim 360$ templates which represent typical M5--M9, L0--L9, and T0--T8 stars. By default, the covered wavelength interval is 0.65--2.55 $\mu$m (from $i_{\rm PS1}$- to $K$-band). To extend the template into the mid-infrared region covered by the WISE data, we calculate the corresponding $W1$ (3.4 $\mu$m) and $W2$ (4.6 $\mu$m) magnitudes, following \cite{2015A&A...574A..78S}, who derived color relations in MLT dwarfs with photometric and spectral data. Following \cite{2017ApJ...849...91M}, we derive the unWISE magnitudes using the synthetic $K$-band magnitude and factors for scaling ($K\_W1$ and $W1\_W2$) for each MLT dwarf template, depending on the spectral type. The following equations were applied:
\begin{equation}
    W1 = K - K\_W1 - 0.783
\end{equation}
\begin{equation}
    W2 = W1 - W1\_W2 - 0.63
\end{equation}

The quasar models were taken from several observed composite spectra which were derived either from low or high redshift quasars spectra. Our first template comes from \cite{2016A&A...585A..87S} who build a luminous quasar composite spectrum at $1 < z < 2$ selected from SDSS. They performed spectroscopic observation with VLT/X-Shooter to cover ultraviolet to near-infrared wavelength simultaneously. This approach ensures that the full rest-frame wavelength range from Ly$\beta$ to 11\,350 \AA~is covered by the composite spectrum with very low contamination from host galaxy stellar emission. 
The second template is from \cite{2016ApJ...833..199J} who utilized 58\,656 Baryon Oscillation Spectroscopic Survey (BOSS) quasar spectra at $2.1 \leq z \leq 3.5$ and created the composite binned by redshift, spectral index, and luminosity. 
The third template that we used was constructed by \cite{2016AJ....151..155H}, which is the averaged spectra of 102\,150 BOSS quasars located at $2.1<z<3.5$.

Three composites of $z>5.6$ quasar spectra were constructed by \cite{2016ApJS..227...11B}. The first one is the averaged spectra of $\sim 100$ sources, the second is built by including only the top 10\% objects by strongest rest-frame Ly$\alpha$+N\textsc{V} equivalent width, and the third was constructed by using the 10\% sources with the smallest Ly$\alpha$+N\textsc{V} EW. These different templates allow us to see how changes due to Ly$\alpha$ emission line strength variation affect the quasar color. 
To reconstruct the intrinsic quasar spectra before absorption of the intervening IGM, we correct each \cite{2016ApJS..227...11B} quasar template using prescription from \cite{2014MNRAS.442.1805I} as calculated at redshift $z=z_\mathrm{median}$ of the quasars used to create the composite. 

Note that the three composite spectra from \cite{2016ApJS..227...11B}, one from \cite{2016ApJ...833..199J}, and one from \cite{2016AJ....151..155H} only cover up to rest-frame wavelength of $\sim 1500$ \AA. So, we extended those templates' coverage by stitching them to the template from \cite{2016A&A...585A..87S} redward of this wavelength. Then, we account for internal reddening by applying the \cite{2000ApJ...533..682C} dust model. Levels of reddening are varied from $E(B-V)=0$ to 0.14 with a 0.02 increment, in addition to two negative reddening of $-0.01$ and $-0.02$ to model the quasars with bluer continuum than covered by the templates.

For completeness, we also make use of the \cite{2014ApJS..212...18B} atlas of galaxy SEDs. These 129 galaxy SED templates include various galaxy types like spirals, ellipticals, and starburst, which are derived from nearby ($z\lesssim0.05$) galaxy observation.
Internal reddening was accounted by applying \cite{2000ApJ...533..682C} dust model to the templates with $A(V)=0$ to 1 with 0.2 increment.
Then, we make a grid of models by shifting all the galaxy templates over the redshift interval $0.0 \leq z \leq 3.0$ with $\Delta z = 0.005$.
However, we only use these templates in visual inspection as final step to make sure that our high-$z$ quasar candidates do not resemble a $z<6$ galaxy SED.

The SED fitting was done by using the EAZY photometric redshift software, created by \cite{2008ApJ...686.1503B}. The way EAZY works is by stepping through a grid of redshifts and trying to find the best template. 
Here we consider the redshift interval of $4.0 \leq z \leq 8.0$ with $\Delta z = 0.003$ for the quasar SED models.
Template spectra are corrected for intervening H\,\textsc{i} cloud absorption following the prescription from \cite{2014MNRAS.442.1805I}. The template fit properly treats flux errors and negative flux measurements because it is done in linear space.
The solutions with smallest reduced-$\chi^2$ ($\chi^2_{\rm red}$) are chosen as best-fit models, which can be computed for each template $i$ as:
\begin{equation}
  \chi^2_{i, \rm red} = \sum_{n=1}^{N} \left(\frac{\textrm{data}_n - f_n(\textrm{model}_i)}{\sigma(\textrm{data}_n)}\right)^2 \bigg/ (N - 1) 
\end{equation}
where the number of photometric data points is $N$ and the degree of freedom is $(N-1)$.

Photometric redshifts ($z_\mathrm{phot}$) may systematically be off from spectroscopic redshifts ($z_\mathrm{spec}$) and we define this as systematic offset bias (see, e.g.\ \citealt{2013MNRAS.432.1483C,2020arXiv200301511N}). This bias can be calculated as $\Delta z = (z_\mathrm{phot} - z_\mathrm{spec})/(1+z_\mathrm{spec})$.
In our quasar candidates selection, we keep track of how many known quasars we can recover at each step (see Table \ref{tab:selection}).
For the sample of 22 known $z>6$ quasars on which we run our SED code on, the average bias is $\langle \Delta z \rangle = 0.01$, scatter is $\sigma_{\Delta z} = 0.02$, and outlier fraction is $|\Delta z| > 0.15 = 0\%$.
Although this represents substantial scatter in the photo-$z$ accuracy, this is already enough to separate between low- and high-$z$ quasars.
Note that our method is most sensitive to select quasars at $6.3\leq z \leq 7.1$ due to our initial color cut criterion (see step 1 in Table \ref{tab:selection}). From a total of $\sim 130$ known quasars in this redshift range, there are only $\sim30$ quasars which were originally discovered within PS1 and DELS data. The other $\gtrsim100$ quasars are found from deeper surveys like SHELLQs, CFHQS, or outside PS1 footprints (e.g.\ DES, VIKING).

\begin{figure*}[htb!]
    \centering
    \plotone{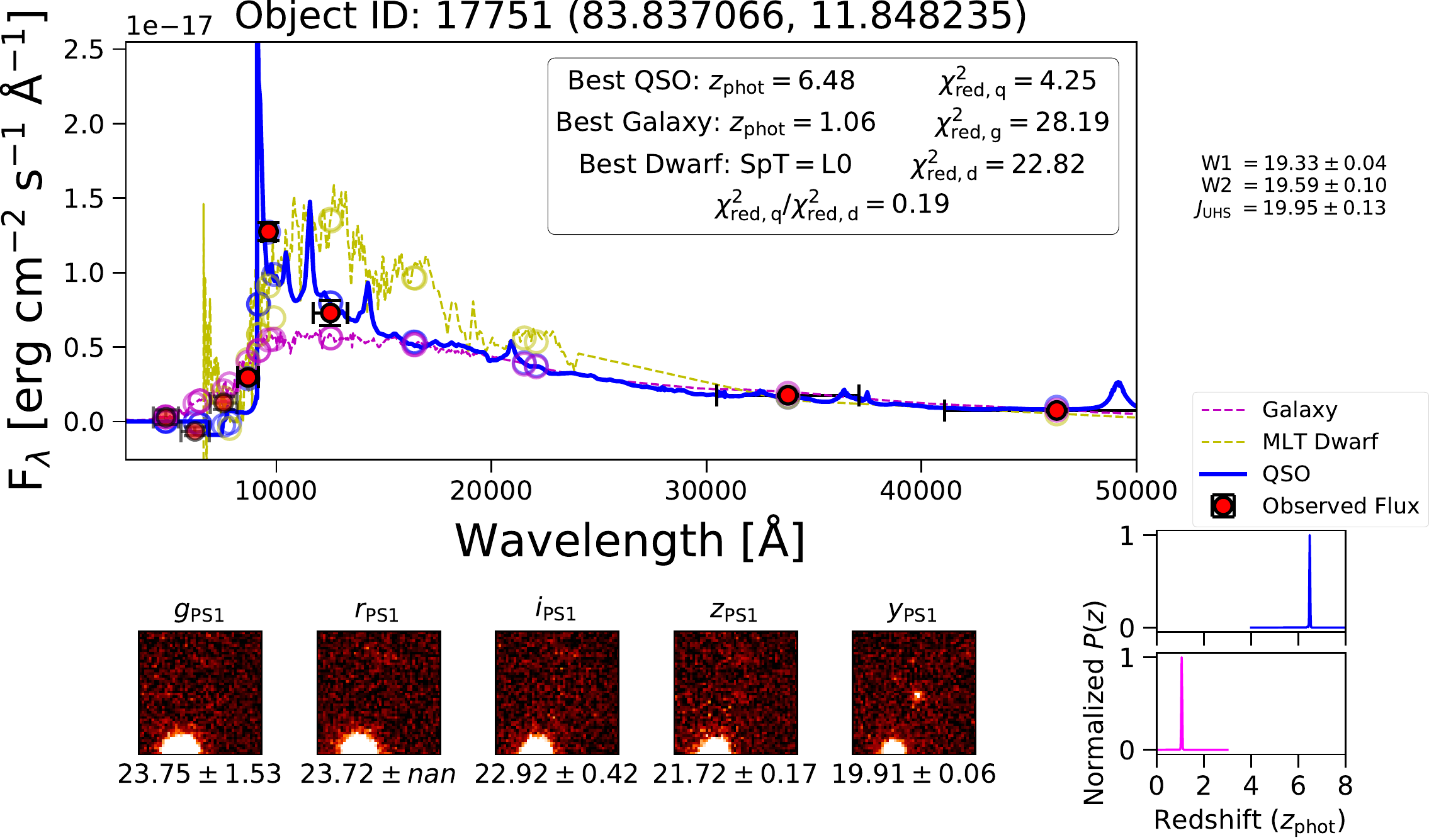}
    \caption{SED fitting result for PSO J083+11. Photometry data is shown with red filled circles with error bars in the top panel. The best-fit quasar spectral template is shown with the blue line and blue circles for model photometry. The same goes for galaxy (magenta) and MLT dwarf models (yellow).
    The bottom panels show 12\arcsec~cutouts in the 5 PS1 bandpasses. All written magnitudes are corrected for Galactic extinction.
    Finally, the bottom right panel shows the probability density function (PDF) of calculated photo-$z$'s for quasar (blue line) and galaxy (magenta line) models.}
    \label{fig:sed_fit}
\end{figure*}

The most probable high-$z$ quasar candidates are selected based on the $\chi^2_{\rm red}$ of the quasar ($\chi^2_{\rm red,q}$) and MLT dwarf ($\chi^2_{\rm red,d}$) model fit, in addition to the estimated photometric redshift (photo-$z$). The ratio of the two $\chi^2_{\rm red}$ ($\chi^2_{\rm red,q}/\chi^2_{\rm red,d}$) is also used as an important factor because this represents how more likely the candidate is a quasar (q) compared to being an MLT dwarf (d). The best values to discriminate between quasar and MLT dwarfs are empirically derived by modeling the SEDs of known PS1 quasars (see the compilation in \citealt{2019arXiv191205614F}) and \cite{2018ApJS..234....1B} MLT dwarfs. An example of SED fitting results can be seen in Figure \ref{fig:sed_fit}.
We choose to use the following criterion:
\begin{equation}
\frac{\chi^2_{\rm red,q}}{\chi^2_{\rm red,d}} < 0.35
\end{equation}
which rejects 89\% of the potential contaminants while recovering 91\% of the known quasars and leaves us with the remaining 7808 candidates.

\edit1{As an additional step, we performed forced photometry on the remaining candidates to confirm the measurements from the catalog and remove spurious detections (see \citealt{2014AJ....148...14B,2016ApJS..227...11B} for reference). For each candidate, the algorithm computed 2\arcsec~radius aperture photometry on the $1\arcmin \times 1\arcmin$ stacked images\footnote{The PS1 stacked images can be retrieved with \texttt{panstamps} (see \url{https://panstamps.readthedocs.io/en/latest/index.html}).} centered on the candidate's coordinate.
It is expected that the aperture photometry is noisier than the PSF photometry. Hence, we imposed a criterion that our own magnitude measurements from the stacked images to be consistent within 2-sigma compared to the magnitudes in the PS1 catalog.
After that, the measured photometry needs to fulfill the Equation~1--5, leaving us with 1263 candidates.}

Finally, visual inspection was done of all the single-epoch and stacked images of PS1 and other public survey, when available. This is a final step to discard non-astronomical sources (e.g.\ CCD artifacts, moving objects, hot pixels). After that, the 155 surviving candidates were included in our list for NIR spectroscopic follow-up observation. A summary of our selection steps can be found in Table \ref{tab:selection}.

\section{Spectroscopic Observations and Analysis} 
\label{sec:spec_obs}

To confirm the quasar nature of our targets, we performed several spectroscopic campaigns. As of July 2020, seven promising candidates have been spectroscopically observed. Five of them are identified as contaminants while the other two, PSO\,J083.8371+11.8482 and PSO J344.1442--02.7664, are previously unknown $z>6$ quasars. The spectroscopically rejected candidates are reported in Appendix \ref{appendix_a} while the discovery of PSO J344.1442--02.7664 is reported in the Appendix \ref{appendix_b}.
We followed up PSO\,J083.8371+11.8482 with deeper near-infrared (NIR) spectroscopy, which we will discuss in the following.

\subsection{Initial classification and redshift determination with Magellan/FIRE} \label{sec:fire_spec}
We firstly confirmed PSO\,J083.8371+11.8482 (hereafter PSO J083+11) as a quasar at $z\sim6.3$ via low-resolution near-infrared (NIR) spectroscopic follow-up by using 6.5m--Magellan/FIRE \citep{2013PASP..125..270S} on December 31, 2018 with total integration time on target of only 5~minutes. The observation was done using the high-throughput prism mode with slit width of 0\farcs6, resulting in a spectral resolution of $R=500$. The observed wavelength range covered by this instrument is $\lambda_{\rm obs}\sim0.82-2.51~\mu$m. 

A second observing run to take a substantially deeper NIR spectrum of PSO J083+11 was done in January and February 2019, using the same telescope. The quasar was observed for 5~hours in the high-resolution echellette mode with the 0\farcs6 slit. This in principle gives us $R = 6000$ spectral resolution or around $\sim50$\,km\,s$^{-1}$ velocity resolution in the wavelength range of 0.82--2.51\,$\mu$m. 
Unfortunately, this second observation run suffered from sub-optimal condition, which degraded the signal-to-noise ratio (S/N).
Still, a quick-look at the spectra showed that all key emission lines -- e.g.\ Ly$\alpha$~$\lambda 1216$, C\,\textsc{iv}~$\lambda 1549$, and Mg\,\textsc{ii}~$\lambda 2798$ -- appear unexpectedly weak. 
We modeled the emission lines and continuum despite the noisy spectrum and found that its continuum power-law slope is consistent with a Type~1 quasar located at $z=6.34$. 
We have to note that the region around Mg\,\textsc{ii} is heavily affected by telluric absorption, which made the accurate line-measurement difficult. To mitigate this issue and to reduce instrument-specific effects, we carried out another spectroscopic observing run.

\subsection{Near-infrared spectroscopy with Gemini/GNIRS} \label{sec:gnirs_spec}
We obtained deep NIR spectroscopy with Gemini Near-InfraRed Spectrograph (GNIRS) at the 8.1m--Gemini North telescope (GN-2019A-FT-204, PI: M.~Onoue) on March 20--22, 2019 with total integration time on target of 8060 seconds. The runs were executed in the cross-dispersed mode to cover the observed wavelength range $\lambda_{\rm obs}\sim0.9-2.5~\mu$m, corresponding to $\lambda_{\rm rest} \sim 1200-3400$ \AA\ in the rest-frame. We utilized the 
`short' camera with a resolution of 0\farcs15 per pixel and a 31.7 l/mm grating.  
We used a slit with aperture of 0\farcs675, resulting in a spectral resolution of $R\sim750$.
The exposure time for single frame was set to 155 seconds and a standard ABBA offset pattern was applied between exposures, to better remove the contribution from the sky emission. The airmass range of the observations was $\sim1.1-1.7$.

\begin{figure*}[htb!]
    \centering
    \plotone{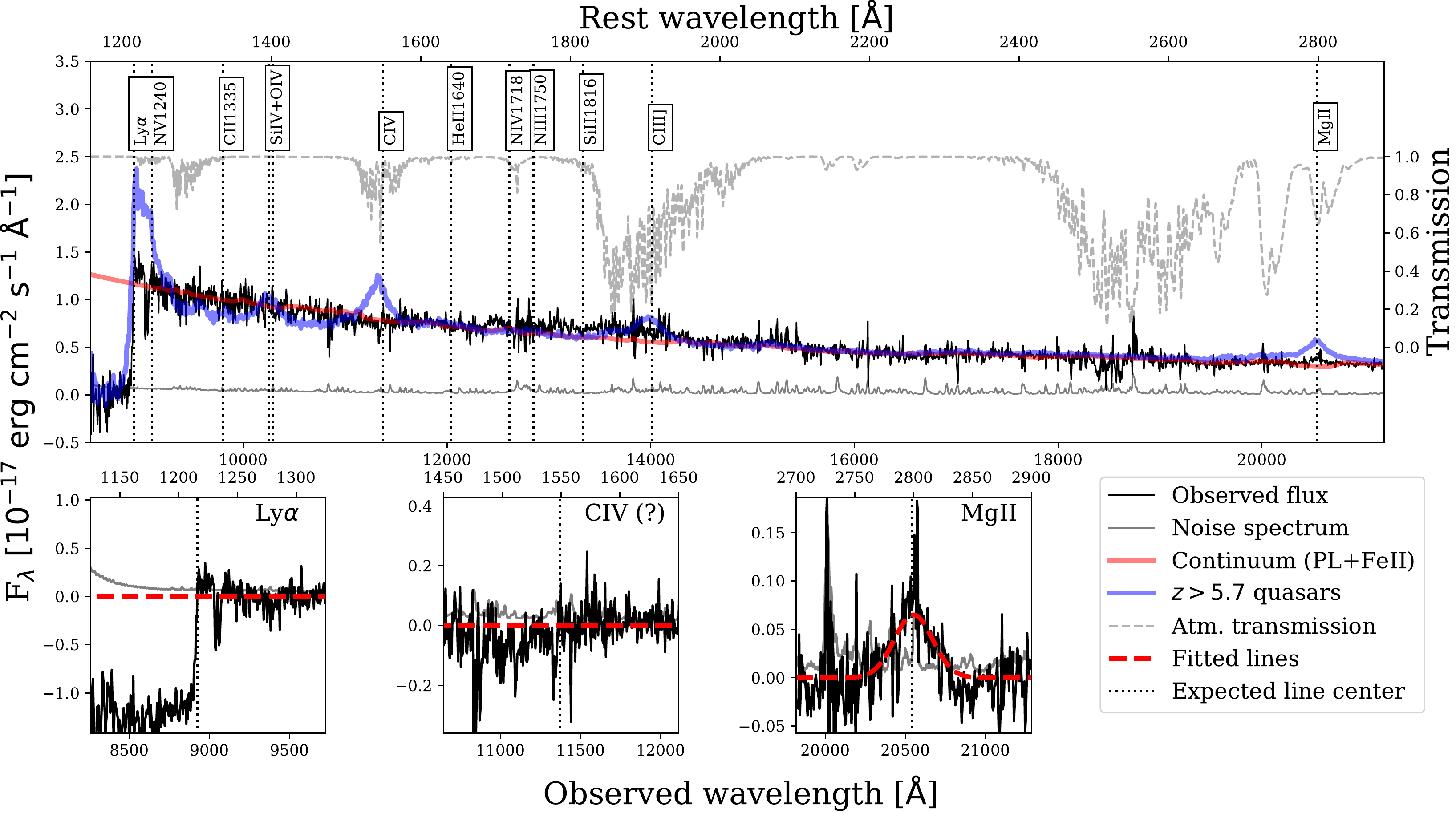}
    \caption{
    \textit{Upper panel}: The spectrum of PSO J083+11 taken with Gemini/GNIRS. The observed flux (black), noise (gray), scaled atmospheric transmission (dashed gray, unit in the right axis), and continuum which is consisted of power-law plus Fe\,\textsc{ii} emission (red) are shown. 
    The expected emission line centers are shown by black vertical dotted lines.
    For comparison, the median composite spectrum of quasars at $z>5.7$ taken from \cite{2019ApJ...873...35S} is plotted in blue. 
    The \cite{2019ApJ...873...35S} composite spectrum is scaled by taking median of PSO J083+11 continuum flux at $\lambda_\mathrm{rest} = 1300-2000$\,\AA~so those two specra are matched with each other.
    \textit{Lower panel:} The continuum subtracted spectrum around the key emission lines (black) and best-fit Gaussian models (dashed red). Note the sharp emission at the center of Mg\,\textsc{ii} is caused by imperfect telluric correction, so it is not real. We also detect a possible metal absorber near the Ly$\alpha$ emission, at $\lambda_\mathrm{obs} = 9037$\,\AA~and 9066\,\AA.}
    \label{fig:spectral_fit}
\end{figure*}

Data reduction was performed with \texttt{PypeIt}\footnote{\url{https://pypeit.readthedocs.io/en/latest/}}, an open source spectroscopic data reduction pipeline \citep{2020arXiv200506505P}.
Each exposure was bias-subtracted and flat-fielded using standard procedures. The wavelength solution was obtained by comparing the spectrum of the sky with the prominent OH \citep{2000A&A...354.1134R} and water lines\footnote{\url{https://hitran.org}}. After removing contamination from cosmic rays, the pipeline optimally subtracts the background by modeling the sky emission with a b-spline function which follows the curvature of the spectrum on the detector. The 1D spectrum of the quasar was created for each exposure using optimal weighting \citep{2003PASP..115..688K}. Relative flux calibration was performed with A-type stars observed before or after the target exposures. We used \texttt{Molecfit}\footnote{\url{https://www.eso.org/sci/software/pipelines/skytools/molecfit}} \citep{2015A&A...576A..78K,2015A&A...576A..77S} to do telluric absorption correction, after which  single-exposure one-dimensional spectra were co-added. All 1D spectra were co-added and scaled to the observed UHS $J$-band photometry ($J_\mathrm{UHS} = 20.09\pm0.13$, not corrected for Galactic extinction) for absolute flux calibration. The reddening due to Galactic extinction is then corrected by using dust map from \citet{2019ApJ...887...93G} and extinction law from \cite{2016ApJ...826..104G}. Figure \ref{fig:spectral_fit} shows the final spectrum in rest frame. From now on, we will use the GNIRS spectrum for the primary analysis instead of the FIRE spectrum, unless otherwise stated explicitly.

\subsection{Modeling the emission lines and underlying continuum}
\label{subsec:spec_model}
To model the PSO J083+11 spectrum, we used a multi-component fitting approach. The global continuum was modelled using combined power-law and UV Fe\,\textsc{ii} templates (\citealt{2001ApJS..134....1V}, \citealt{2006ApJ...650...57T}, \citealt{2007ApJ...662..131S}). Free parameters are the scaling factor for each component and the power-law continuum slope.
We implemented this approach by using a modified version of \texttt{PyQSOFit}\footnote{\url{https://github.com/legolason/PyQSOFit}}, a code to fit typical quasar spectra \citep{2018ascl.soft09008G}.
We improved the performance of \texttt{PyQSOFit} with respect to high-$z$ quasars by optimizing the fit of the narrow and broad emission lines, substituting the continuum windows and other minor modifications.
We define the continuum window for the fit by iteratively marking emission- and telluric-line-free regions. The following wavelength ranges were selected as continuum windows: $\lambda_\mathrm{rest}=$~1285--1290\,\AA, 1315--1325\,\AA, 1350--1370\,\AA, 1445--1465\,\AA, 1580--1650\,\AA, 2140--2300\,\AA, 2340--2400\,\AA, 2420--2480\,\AA, 2630--2710\,\AA, 2745--2765\,\AA, and 2850--3000\,\AA.
\begin{deluxetable}{lrr}[htb!]
	\tablenum{2}
	\tablecaption{Derived physical parameters of PSO J083+11.}
	\label{tab:phys_par}
	\tablehead{
		\colhead{Parameter} & \colhead{Value} & \colhead{Unit}
	}	
	\startdata
	$\alpha_\lambda$ & $-1.66^{+0.01}_{-0.04}$ \\
	$M_{1450}$ & $-26.67 \pm 0.01$ & mag \\
    $M_{\rm BH}$ & $\left(2.00^{+0.74}_{-0.44}\right) \times 10^9$ & $M_\odot$\\
    $L_{\rm bol}$ & $\left(1.33^{+0.01}_{-0.03}\right) \times 10^{47}$ & erg\,s$^{-1}$\\
    $L_{\rm bol}/L_{\rm Edd}$ & $0.51^{+0.13}_{-0.17}$ & \\
    FWHM (Mg\,\textsc{ii}) & $4140^{+880}_{-430}$ & km s$^{-1}$ \\
    $\Delta v (\rm Mg\,\textsc{ii}-[C\,\textsc{ii}])$\tablenotemark{a} & $237 \pm 150$ & km~s$^{-1}$ \\
    EW$(\rm Mg\,\textsc{ii})_{rest}$ & $8.71^{+0.67}_{-0.64}$ & \AA \\
    EW$(\rm Ly\alpha+N\,\textsc{v})_{rest}$ & $5.65^{+0.72}_{-0.66}$ & \AA \\
    EW$(\rm C\,\textsc{iv})_{rest}$\tablenotemark{b} & $\leq 5.83$ & \AA \\
    \hline
    $R_p$ & $1.17 \pm 0.32$ & pMpc \\
    $t_\mathrm{Q}$ & $10^{3.4\pm0.7}$ & yr \\
    \hline
    $z_{\rm [C\,\textsc{ii}]}$ & $6.3401 \pm 0.0004$ & \\
    FWHM ([C\,\textsc{ii}]) & $229 \pm 5$ & km s$^{-1}$ \\
    Flux ([C\,\textsc{ii}]) & $10.22 \pm 0.35$ & Jy km s$^{-1}$ \\
    $S_{244\,\mathrm{GHz}}$ & $5.10 \pm 0.15$ & mJy \\
    $S_{258\,\mathrm{GHz}}$ & $5.54 \pm 0.16$ & mJy \\
    \hline
    $L_{\rm [C\,\textsc{ii}]}$ &  $(1.04 \pm 0.04) \times 10^{10}$ & $\,L_\odot$ \\
    $L_{\rm FIR}$ &  $(1.22 \pm 0.07) \times 10^{13}$ & $\,L_\odot$ \\
    $L_{\rm TIR}$ &  $(1.72 \pm 0.09) \times 10^{13}$ & $\,L_\odot$ \\
    SFR$_{\rm [C\,\textsc{ii}]}$ & $800-4900$ & $\,M_\odot\,\mathrm{yr^{-1}}$ \\
    SFR$_{\rm TIR}$ & $900-7600$ & $\,M_\odot\,\mathrm{yr^{-1}}$ \\
    $M_{\rm dust}$ & $(4.88 \pm 0.14) \times 10^8$ & $M_\odot$\\
	\enddata
	\tablenotetext{a}{Velocity shift of Mg\,\textsc{ii} with respect to [C\,\textsc{ii}].}
	\tablenotetext{b}{This is the 3-sigma upper limit value, see Section \ref{sec:weak_line}.}
\end{deluxetable}

The rest-frame 1450\,\AA~absolute magnitude and the monochromatic luminosity at 3000\,\AA~($L_\lambda (3000\,\text{\AA})$) were calculated from the best-fit power-law continuum. Then, the bolometric luminosity $L_{\rm bol}$ was derived using the empirical correction from \cite{2006ApJS..166..470R}:
\begin{equation}
    L_{\mathrm{bol}} = 5.15 \times \lambda L_{3000}
    \label{eq:lbol}
\end{equation}
After subtracting the best-fit continuum and scaled iron templates, each broad emission line was modeled with single Gaussian functions. We applied Monte Carlo simulation and created mock spectra to estimate the errors of the derived parameters.
Random flux density errors are drawn assuming a normal distribution using the noise spectrum, then applied to the raw spectrum with 1000 iterations.
The measurement lower and upper limit values are taken as the 16th and 84th percentiles of the distribution of these repeated measurements, respectively.
Finally, we obtained the equivalent width (EW), central wavelength, full width at half-maximum (FWHM), and velocity dispersion for each line. 

However, as seen in Figure \ref{fig:spectral_fit}, the only strongly detected broad line is Mg\,\textsc{ii}, where we found FWHM~(Mg\,\textsc{ii})~$=4140^{+880}_{-430}$~km~s$^{-1}$ and EW~(Mg\,\textsc{ii})$_\mathrm{rest}$~$=8.71^{+0.67}_{-0.64}$~\AA. 
Moreover, we obtained the redshift of $z_{\rm Mg\,\textsc{ii}} = 6.346 \pm 0.001$, 
determined from the observed Mg\,\textsc{ii} central wavelength.
The Ly$\alpha$ is weak and we do not significantly detect the C\,\textsc{iv} line. Derivation of Ly$\alpha$ and C\,\textsc{iv} EWs will be explained later in Section \ref{sec:weak_line}. Assuming that C\,\textsc{iv} will have a FWHM similar or larger than that of Mg\,\textsc{ii}, we do not see any potential broad absorption line signatures in the region where C\,\textsc{iv} is expected to be present (see bottom panel of Figure \ref{fig:spectral_fit}). Hence we conclude that the absence of C\,\textsc{iv} is not simply caused by broad absorption line phenomenon, but rather than actual nature of this quasar (see Section \ref{sec:weak_line}).

\subsection{Black hole mass and Eddington ratio} 
\label{sec:bh_properties}

The mass of the black hole is derived from our single-epoch NIR spectra. With the assumption that the virial theorem is valid for the BLR dynamics, we use the scaling relation for the Mg\,\textsc{ii} line from \cite{2009ApJ...699..800V}:
\begin{equation}
    \frac{M_{\mathrm{BH}}}{M_\odot} = 10^{6.86} \bigg(\mathrm{\frac{FWHM\,(Mg\,\textsc{ii})}{10^3\,km\,s^{-1}}}\bigg)^2 \bigg(\frac{\lambda L_\lambda (3000\,\text{\AA})}{10^{44} \, \mathrm{erg\,s^{-1}}}\bigg)^{0.5}
    \label{eq:bh_mass}
\end{equation}
where $\lambda L_\lambda (3000\,\text{\AA})$ is the rest-frame luminosity at 3000 \AA\ and FWHM\,(Mg\,\textsc{ii}) is the Mg\,\textsc{ii} full width at half maximum.
Then, we calculated the Eddington luminosity as:
\begin{equation}
    L_\mathrm{Edd} = 1.3 \times 10^{38} \bigg(\frac{M_{\mathrm{BH}}}{M_\odot}\bigg)\,\rm erg\,s^{-1}
\end{equation}
After that, we can derive the Eddington ratio as $L_{\rm bol}/L_{\rm Edd}$. 
We obtained a black hole mass of $\log (M_{\rm BH}/M_\odot) = 9.30^{+0.16}_{-0.10}$ and normalized accretion rate of $L_{\rm bol}/L_{\rm Edd} = 0.51^{+0.13}_{-0.17}$ for the GNIRS spectrum. As a comparison, we derived $\log (M_{\rm BH}/M_\odot) = 9.06^{+0.24}_{-0.16}$ and $L_{\rm bol}/L_{\rm Edd} = 0.77_{-0.33}^{+0.33}$ from the combined FIRE spectra.
We report the error in measured virial black hole mass which we calculated by propagating monochromatic luminosity errors and Mg\,\textsc{ii} line width.
However, please note that we did not explicitly add the systematic errors which tend to be larger ($\approx 0.5$~dex) compared to random measurement errors \citep{2009ApJ...699..800V,2013BASI...41...61S}.

\begin{figure}[htb!]
    \centering
    \epsscale{1.1}
    \plotone{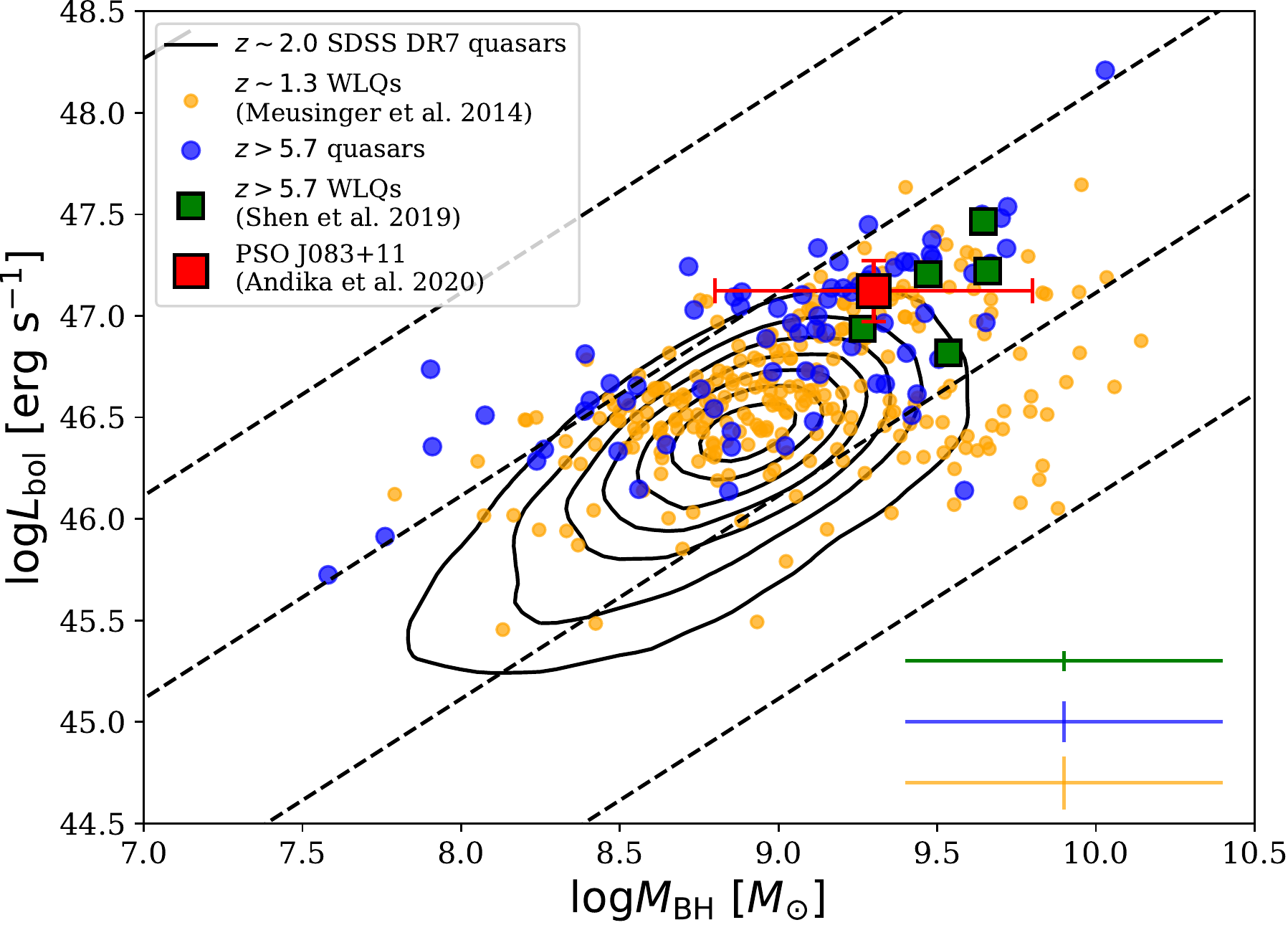}
    \caption{The BH mass--bolometric luminosity plane of quasars at various redshifts. 
    Those parameters are calculated for sample of $z\sim1.3$ WLQs (magenta; \citealt{2014A&A...568A.114M}), $z>5.7$ WLQs (green; \citealt{2019ApJ...873...35S}), and other $z>5.7$ quasars compiled from literature (blue; see text for details).
    The 0.5 dex systematic uncertainty associated with the Mg\,\textsc{ii}-based BH mass estimate \citep{2013BASI...41...61S} as well as the measurement errors are taken into account in the error bars (red square with error bars). Contours show the distribution of the $z\sim2$ SDSS DR7 quasars \citep{2011ApJS..194...45S}. 
    The diagonal lines show Eddington ratios of $L_{\rm bol}/L_{\rm Edd}=$~10, 1, 0.1, and 0.01 from top left to bottom right.
    \edit1{The associated typical uncertainties for each sample are shown as error bars in the bottom right.}
    }
    \label{fig:masslum}
\end{figure}

In order to compare PSO J083+11 properties to other quasars at high redshifts, we compile Mg\,\textsc{ii} line width and continuum luminosity measurements for $\sim 70$ previously published $z>5.7$ quasars \citep{2007AJ....134.1150J,2010AJ....139..906W,2011ApJ...739...56D,2014ApJ...790..145D,2015Natur.518..512W,2017ApJ...849...91M,2018ApJ...867...30E,2019MNRAS.484.2575T,2019ApJ...870L..11F,2019ApJ...873...35S,2019ApJ...880...77O}. 
Then, we classify 5 quasars with EW$(\rm C\,\textsc{iv}) < 10$\,\AA~in \cite{2019ApJ...873...35S} as high-$z$ weak emission line quasars (WLQs). On the other hand, 261 WLQs at $z\sim1.3$ are taken from \cite{2014A&A...568A.114M}.
We use their Mg\,\textsc{ii} line width and continuum measurements to derive bolometric luminosity and virial mass with the same cosmology assumption and scaling relation (see Equation \ref{eq:lbol} and \ref{eq:bh_mass}) as those applied for PSO J083+11. Figure \ref{fig:masslum} shows the distribution of calculated parameters in the BH mass--bolometric luminosity plane.
Our object populate the same $L_{\rm bol}/L_{\rm Edd}$ and $M_{\rm BH}$ parameter space as other quasars at $z>5.7$ and those observed at $z\sim2$ from the SDSS Data Release 7.
Hence, we find that PSO J083+11 is powered by a typical matured and actively accreting SMBH, which has been reported in other quasars at similar luminosity ranges. A summary of calculated physical parameters is shown in Table \ref{tab:phys_par}.

\section{Searching for a Lensing Galaxy} 
\label{sec:hst}
It is well known that gravitational lensing can potentially boost a quasar's observed flux which could lead to a substantial overestimation of black hole masses powering. This would have an impact on our understanding of a high-$z$ quasar's  intrinsic properties. 
Furthermore, a large intrinsic lensing fraction among luminous high-$z$ quasars has been predicted \citep{2002ApJ...580...63C,2002ApJ...581..886W}. Moreover, \cite{2019ApJ...870L..12P} predicted that there should be many mildly magnified ($\mu \leq 10$) quasars at $z>6$ with extremely small image separation ($\Delta \theta \lesssim0\farcs2$). The first confirmation of this prediction is the lensed quasar at $z=6.51$ found by \cite{2019ApJ...870L..11F}, J043947.08+163415.7.
While there is no obvious companion in the vicinity of PSO J083+11 as judged from the discovery images, the ground-based seeing prohibits detecting various potential lens configurations with small separations. Hence, we have used the Hubble Space Telescope (HST) to test the lensing hypothesis for this quasar.

\subsection{Near-infrared imaging with HST}
We obtained high resolution imaging for PSO J083+11 by using HST (GO 15707, PI: K. Jahnke). 
Our goal is to test the quasar image being subject of gravitational lensing by utilizing two methods. The first one is by searching for multiple quasar images using the NIR F125W filter ($\lambda_\mathrm{eff}=12365$\,\AA) on the Wide Field Camera 3 (WFC3). The second one is directly searching for an intervening galaxy with the Advanced Camera for Surveys (ACS) ramp-filter FR853N ($\lambda_\mathrm{eff}=8528$\,\AA) in the quasar Gunn-Peterson absorption trough just short-ward of Ly$\alpha$, where the quasar light will be nearly fully absorbed, maximizing visibility of any intervening lensing galaxy.
Each band was exposed for two orbits, for a total integration time of $\sim80$ minutes per filter. For the  WFC3/IR imaging, we rotated the field between the two orbits by $\sim 15\deg$, to analyze not only the images directly, but also the difference image that reduced the interference of instrumental point-source spikes integral to the HST point spread function. Data reduction was carried out with the HST pipeline for a final image pixel scale is 0\farcs128 pixel$^{-1}$. The 5$\sigma$ surface brightness limit for a 1~arcsec$^2$ aperture is $\sim26$~mag~arcsec$^{-2}$.

\begin{figure*}[htb!]
    \centering
    \plotone{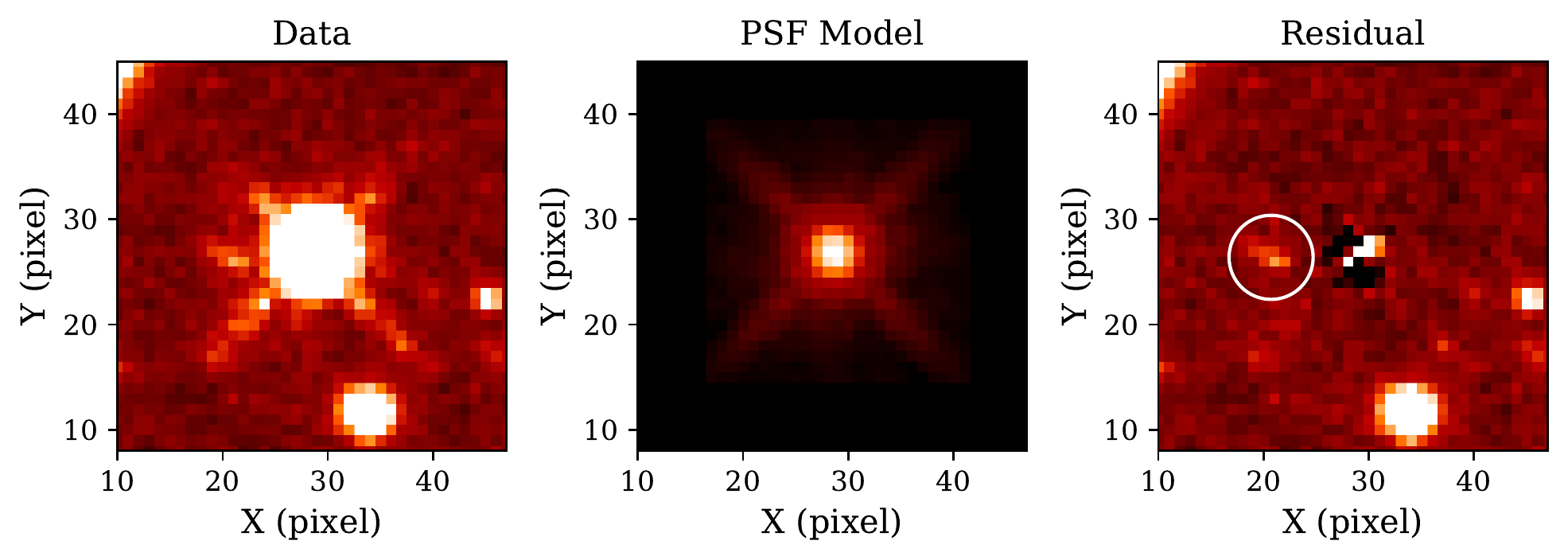}
    \caption{HST WFC3/IR F125W imaging of PSO J083+11. The observed quasar emission (left), PSF model (middle), and the PSF-subtracted image (right) are shown. The image pixel scale is 0\farcs128 pixel$^{-1}$. The white circle in the residual image marks the aperture used to determine the aperture photometry of the (likely) foreground galaxy, located at $\sim1$\arcsec\ from the central quasar. We use a 4-pixel radius aperture (equivalent to $\sim 0\farcs5$) to get a F125W band $\mathrm{magnitude} = 25.42 \pm 0.07$.}
    \label{fig:psf_subtraction}
\end{figure*}

Our next step is to analyze the WFC3/F125W HST image and search for extended emission, which could be due to an intervening lensing galaxy, companion source, or the host galaxy of PSO J083+11 itself. Our approach is to remove the point-like emission from the central quasar, utilizing the \texttt{Photutils}\footnote{\url{https://photutils.readthedocs.io/en/stable/}} software \citep{Bradley_2019_2533376}.
We chose eight stars in the field to construct a point-spread function (PSF) using median-averaging. The stars are chosen so they are sufficiently away from the edge of the CCD or potential contaminants. They also have to be $1.5-15\times$ brighter than the quasar, to get accurate PSF wings.
Note that we did not take into account the effect of the spectral types of selected reference stars, hence there might be color dependent uncertainties due to  systematic SED differences.
Then, this model is fitted to the quasar's nuclear emission in the image, allowing the PSF centroid to move by less than a pixel.
The observed image and PSF subtraction residual are shown in Figure \ref{fig:psf_subtraction}.

\subsection{Modeling a possible gravitational lensing effect} \label{sec:hst_result}
We identified emission in the WFC3/F125W images which could be attributed to an intervening foreground galaxy having a potential gravitational lensing effect, located at 1\arcsec\ to the southwest from the central quasar. However, we could not constrain a redshift of this emission by using HST data only. Its AB-magnitude was estimated using aperture photometry, with an aperture of 4-pixel ($\sim 0\farcs5$) radius with a value of $\mathrm{mag_{F125W}} = 25.42 \pm 0.07$.
However, this emission doesn't show up in the ACS/FR853N image. We did the photometry with the same aperture size and obtained 5-sigma upper limit magnitude of $\mathrm{mag_{FR853N}} = 23.25$.

We test whether the intervening galaxy could potentially boost the apparent quasar emission by modeling the galaxy's emission from the far-ultraviolet to the microwave regimes using \texttt{BAGPIPES}\footnote{\url{https://github.com/ACCarnall/bagpipes}} \citep{2018MNRAS.480.4379C}.
We used the measured flux from WFC3/F125W image aperture photometry, complemented with upper limit fluxes from ACS/FR853N and PS1 bands as input data.
Here, we assume the foreground emission which was found in the F125W images is from a star-forming main sequence galaxy.
To derive the stellar mass and star formation rate (SFR), we need to scale the galaxy SED by choosing a particular star formation history (SFH) model. The simplest form of a parametric model uses up to three shape parameters and a normalization.
The most simple and widely used parametric SFH model is exponentially declining (tau model, $\tau$).
For this work, we consider the delayed exponentially declining SFHs, which is a tau model multiplied by the time since star formation began ($T_0$). This would remove both a discontinuity in SFR at $T_0$ and models with rising SFHs if $\tau$ is large (see \citealt{2018MNRAS.480.4379C}).
The model will fit the total formed stellar mass and time since star formation began with uniform priors in the value ranges of $M_{\rm formed} = 10^{1-15} M_\odot$ and $T_0 = 0.5-0.8$ Gyr, respectively. 
We assume a fixed value for the SFR timescale ($\tau = 0.3$ Gyr), for the metallicity ($Z=0.02$; equals to solar metallicity), and for the nebular emission parameter ($\log(U)=-3$).
The value of $A_V=1.0$ mag was also applied following the \cite{2000ApJ...533..682C} extinction law.

\begin{figure}[htb!]
    \centering
    \epsscale{1.1}
    \plotone{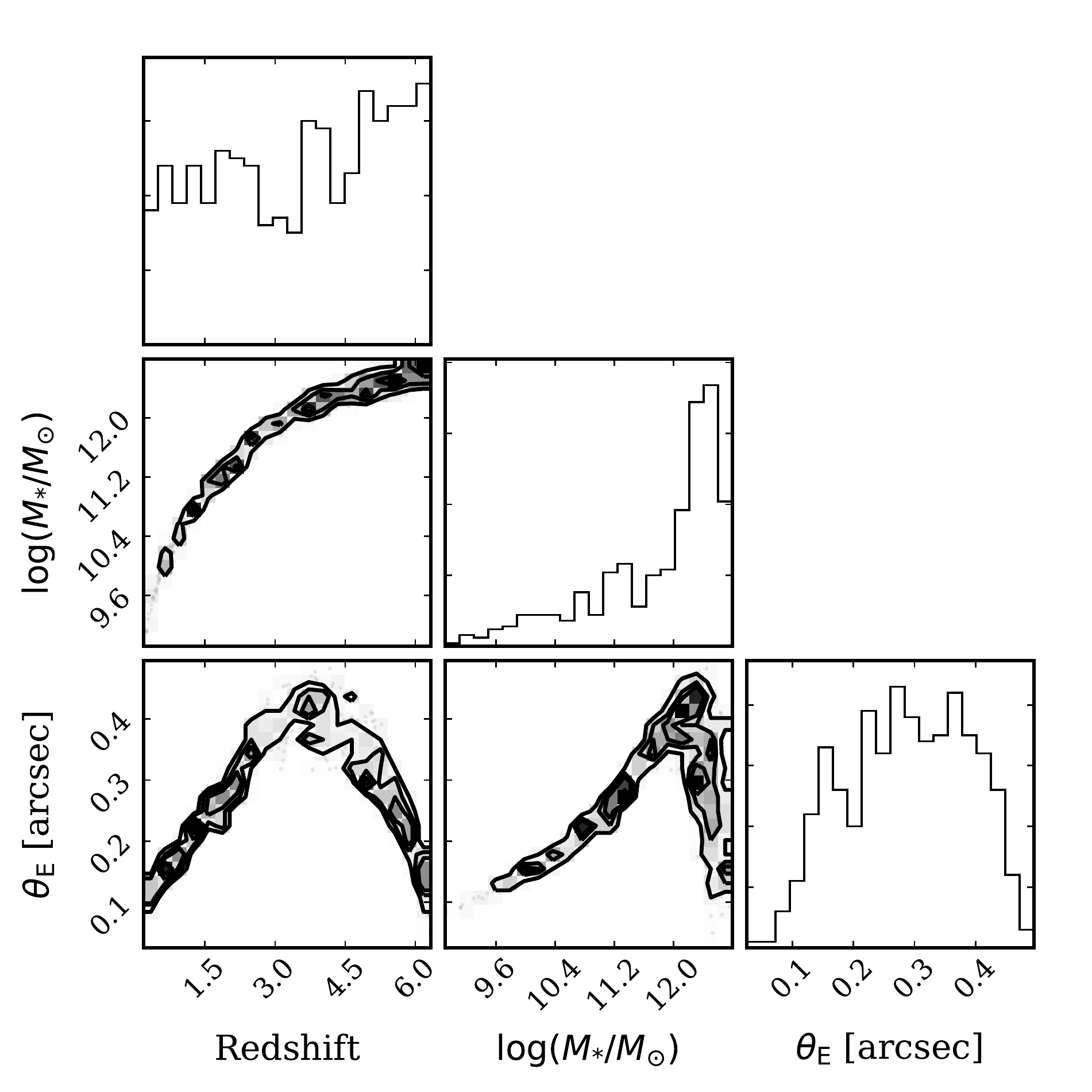}
    \caption{
        A corner plot showing the calculated Einstein angles as a function of simulated galaxy masses and redshifts.
        }
    \label{fig:lens_model}
\end{figure}
For us, the central output calculated by \texttt{BAGPIPES} is the redshift-dependent galaxy stellar mass, which we will use to  constrain the possible magnification for PSO J083+11 by applying the lensing equation (see \citealt{2006eac..book.....S} for a review).
Given the upper limit for the mass-to-light ratio ($\sim100$) for disk and ellipticals, we can estimate the maximum total lens mass, including the dark matter contribution, from the galaxy stellar mass output of the model.
Assuming a point-mass gravitational lens configuration, the possible combinations of Einstein angle ($\theta_{\rm E}$) and magnification ($\mu$) can be calculated. Given the calculated lensing galaxy masses we find limits of $\theta_{\rm E} \leq 0\farcs5$ and $\mu \leq 1.07$. Therefore, strong magnification of the quasar’s emission by the foreground galaxy can be excluded.
The calculated Einstein angles as a function of simulated galaxy masses and redshifts is shown in Figure~\ref{fig:lens_model}.

\section{Weak emission lines and young quasar accretion lifetime} 
\label{sec:weak_line}

\begin{figure}[htb!]
    \centering
    \epsscale{1.1}
    \plotone{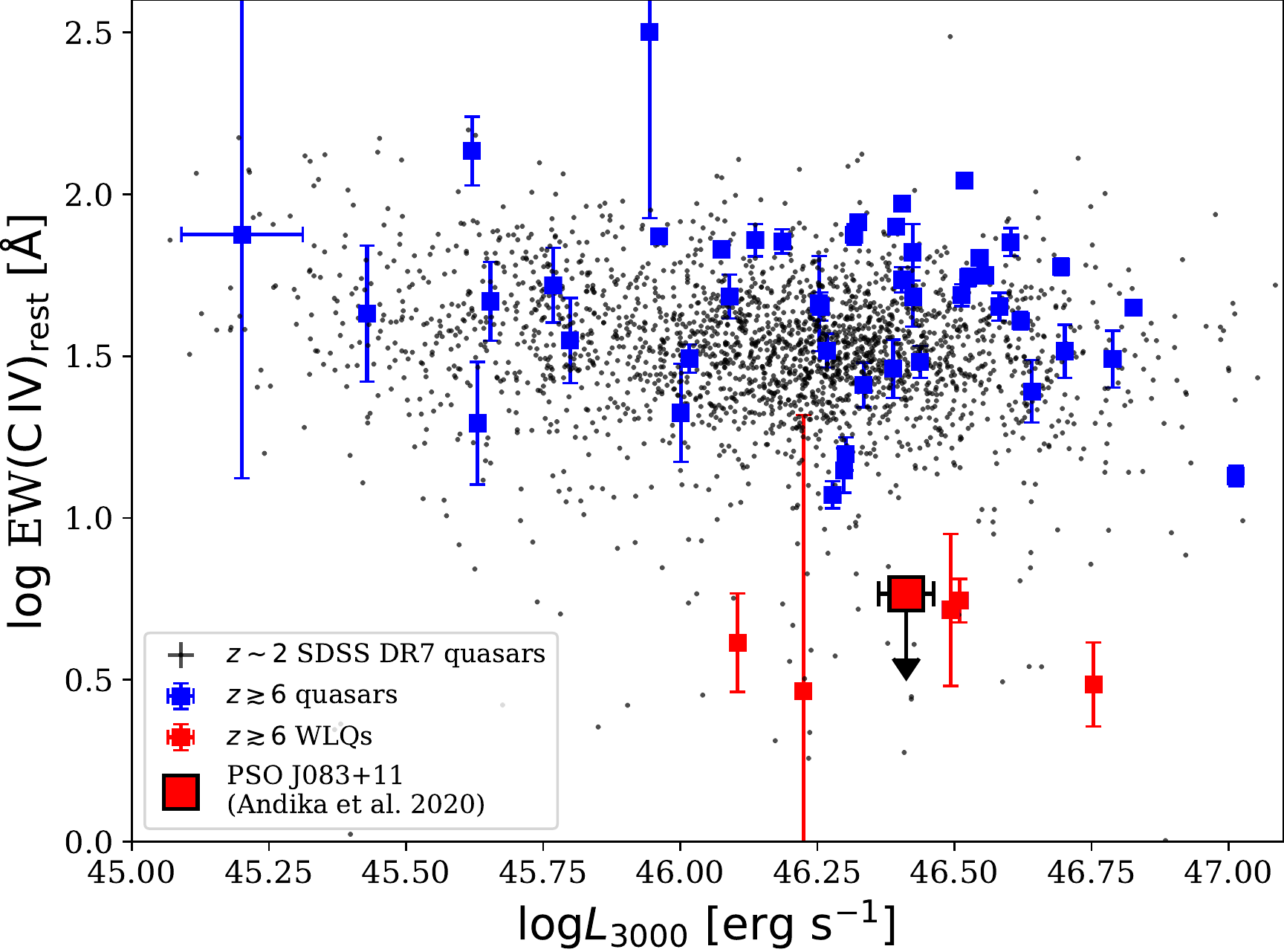}
    \caption{The C\,\textsc{iv} rest-frame EW as a function of continuum luminosity at 3000\,\AA. The black dots show sample of $z\sim2$ quasars from \cite{2011ApJS..194...45S}. The $z>5.7$ normal quasars (blue squares) and WLQs (red squares) which are taken from \cite{2019ApJ...873...35S} are also shown. It is clear that PSO J083+11 is the $3\sigma$ outlier at the low-end of normal quasars' EW$(\rm C\,\textsc{iv})_{rest}$ log-normal distribution.}
    \label{fig:civ_l3000}
\end{figure}

As already noted in Section \ref{subsec:spec_model}, the quasar broad emission lines are very weak or absent. 
We follow the prescription by \citet{2009ApJ...699..782D} to determine the Ly$\alpha$+N\,\textsc{v} equivalent width, where the fluxes above the power-law continuum in the range of 1160--1290 \AA~were integrated. 
This region is dominated by blended Ly$\alpha$~$\lambda 1216$ and N\,\textsc{v}~$\lambda 1240$ components, although there is also a small fraction of S\,\textsc{iii}~$\lambda 1263$.
The calculated Ly$\alpha$+N\,\textsc{v} equivalent width is EW$(\rm Ly\alpha+N\,\textsc{v})_{rest} = 5.65^{+0.72}_{-0.66}$~\AA. In addition, we applied the same procedure for the wavelength range of 1500--1600 \AA\ to estimate the 3-sigma upper limit of the C\,\textsc{iv}~$\lambda 1549$ equivalent width, for which we find EW$(\rm C\,\textsc{iv})_{rest} \leq 5.83$~\AA. From these measurements, we can classify PSO J083+11 as a weak-line quasar according to the empirical definition of \cite{2009ApJ...699..782D}. 
The comparison between normal quasars and WLQs' EW$(\rm C\,\textsc{iv})_{rest}$ as a function of continuum luminosity is shown in Figure \ref{fig:civ_l3000}.
Although originally WLQs are defined as quasars having EW$(\rm C\,\textsc{iv})_{rest} < 10$ and found in low redshift (``low-$z$'', $z\lesssim5$) quasars, this definition is still valid for high-$z$ quasars. \cite{2019ApJ...873...35S} showed that there is no significant redshift evolution of EW$(\rm C\,\textsc{iv})_{rest}$, at least up to $z\sim6$ (see their Figure 8).

\cite{2017ApJ...840...24E} showed that the lifetime of quasars can be inferred from their proximity zones size. By definition, the proximity zone is the region of enhanced Ly$\alpha$ forest transmission close to the quasar resulting from its own ionizing radiation. The IGM will have a finite response time to reach a new ionization equilibrium state due to the quasars' radiation with a timescale of $t_{\rm eq} \approx \Gamma^{-1}_{\rm H\,\textsc{i}} \approx 3 \times 10^4$\,yr. Here, $\Gamma_{\rm H\,\textsc{i}}$ is the rate of photoionization \citep{2017ApJ...840...24E}.

\begin{figure}[htb!]
    \centering
    \epsscale{1.1}
    \plotone{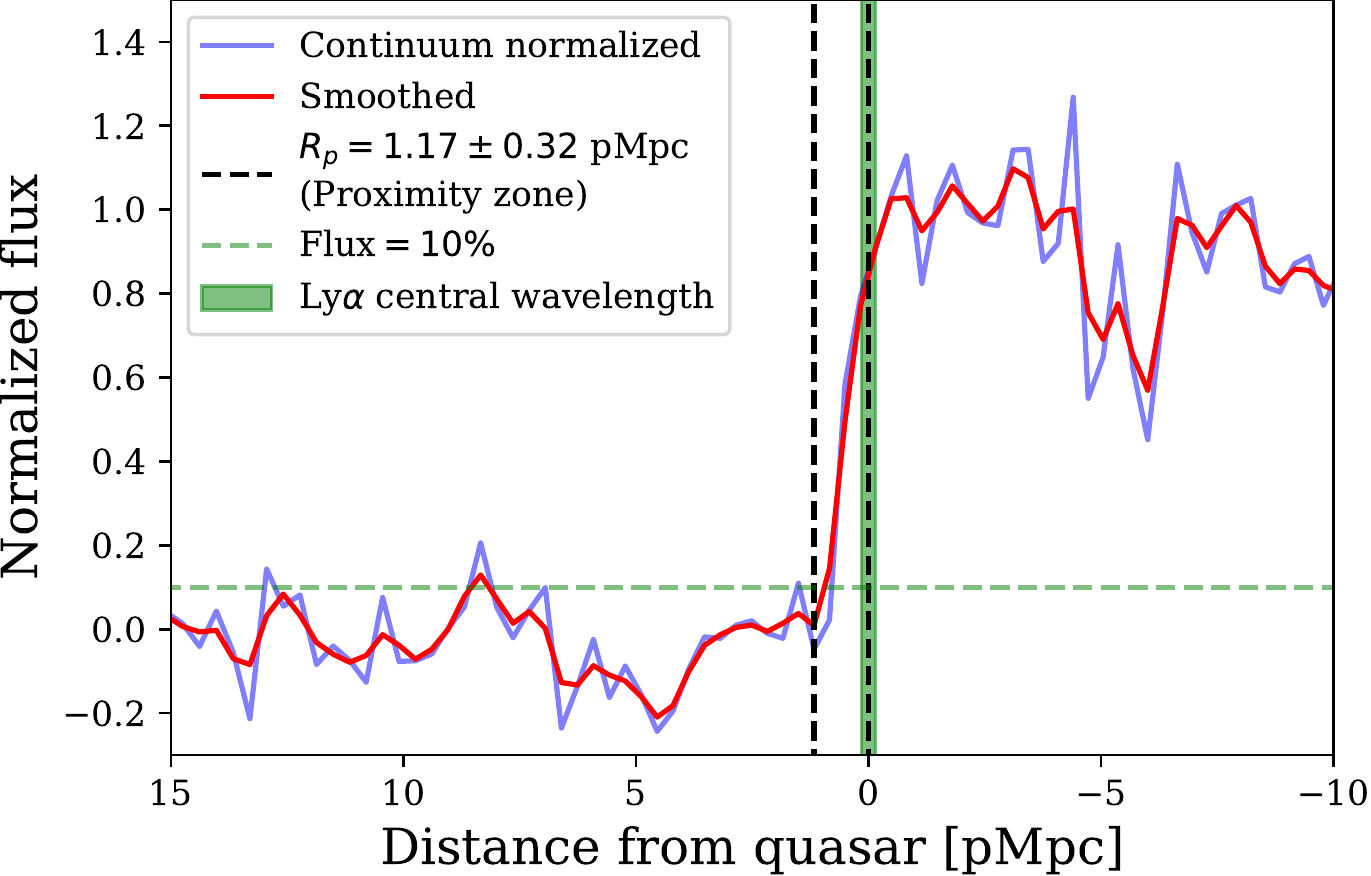}
    \caption{Measured proximity zone size ($R_p$) of PSO J083+11, marked with two black dashed-lines. The expected position of Ly$\alpha$ central wavelength is marked with green line. Continuum normalized and smoothed spectra are shown as blue and red lines, respectively.
    Here we observe a rather small proximity zone of $R_{\mathrm p}=1.17\pm0.32$ pMpc.}
    \label{fig:proximity_zone}
\end{figure}
\begin{figure*}[htb!]
    \centering
    \epsscale{1.1}
    \plotone{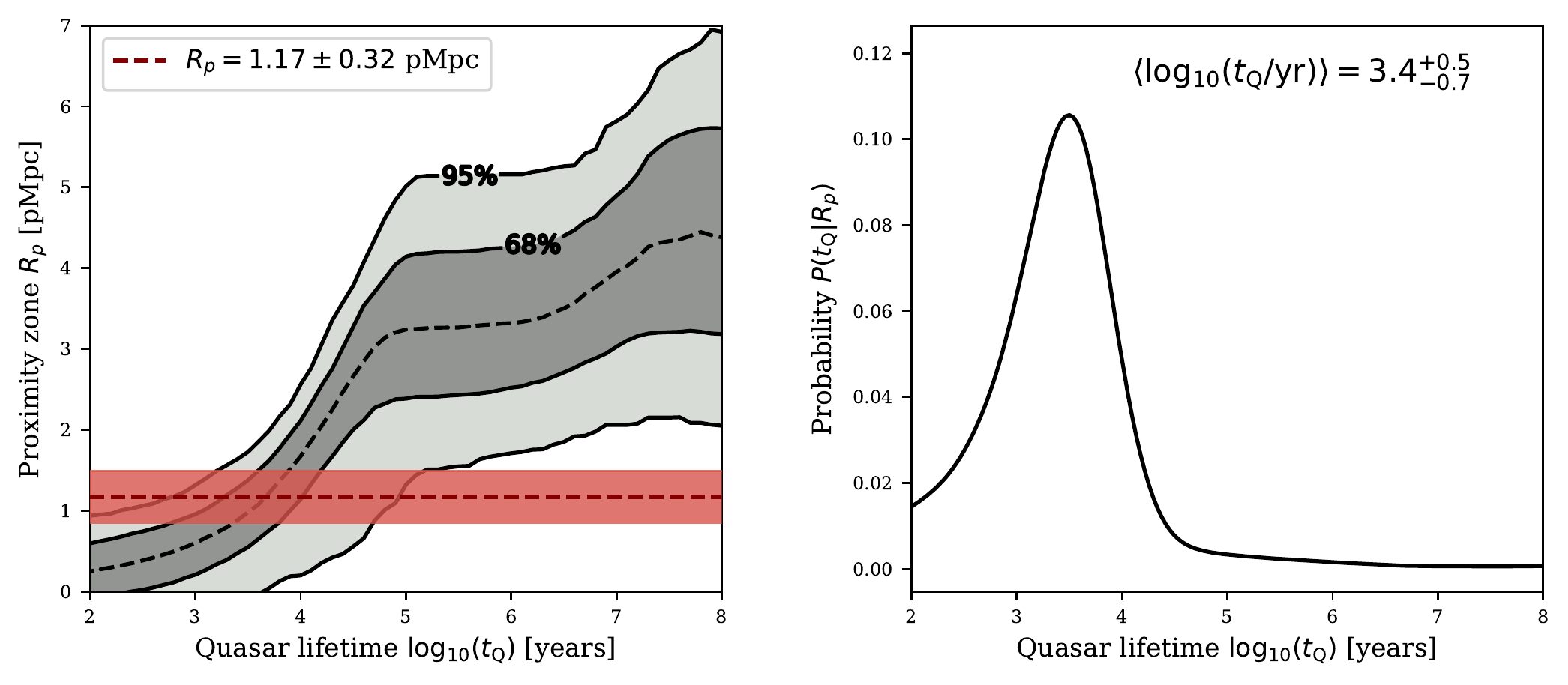}
    \caption{\textit{Left panel}: dependence of proximity zone size $R_p$ on quasar lifetime ($t_\mathrm{Q}$), adopted from literature \citep{2016MNRAS.457.3006D,2017ApJ...840...24E} who have done radiative transfer simulations for quasars with similar absolute magnitude ($M_{1450} = 26.67$) and redshift ($z=6.3401$) as PSO J083+11.
    The $R_p = 1.17 \pm 0.32$ is indicated by red lines, 
    \textit{Right panel}: probability distribution of $t_\mathrm{Q}$ which shows that PSO J083+11 is a young object with $t_\mathrm{Q} = 10^{3.4\pm0.7}$ years (95\% confidence interval).
    }
    \label{fig:quasar_age}
\end{figure*}

In practice, we continuum-normalized the quasar Magellan/FIRE spectrum and applied a convolution with a 20\,\AA~resolution boxcar kernel. Then, proximity zone size is defined as the distance from center of Ly$\alpha$ to shorter wavelengths where the transmitted flux first drops below 10\% of the level at the line \citep{2006AJ....132..117F}. Note that this is done in the observed-frame of wavelength. The result is shown in Figure \ref{fig:proximity_zone}.
Conversion of proximity zone size to quasar lifetime can be inferred from radiative transfer simulations \citep{2016MNRAS.457.3006D} for quasars with similar redshift and luminosity as our quasar, and is shown in the Figure \ref{fig:quasar_age}. 
We observe a rather small proximity zone ($R_{\mathrm p}=1.17\pm0.32$ pMpc) in the spectrum which implies that PSO J083+11 has lifetime of only $t_\mathrm{Q} = 10^{3.4\pm0.7}$ years. As a comparison, the typical proximity zone sizes of quasars with $-27.5\lesssim M_{1450}\lesssim-26.5$ are $R_{\mathrm p}=3-7$ pMpc while their typical lifetimes are $t_\mathrm{Q} = 10^{5-6}$ years \citep{2017ApJ...840...24E}.

\section{Probing the Quasar Host Galaxy} 
\label{sec:alma}

\subsection{Sub-millimeter observation with ALMA}
\label{sec:alma_obs}

The ALMA band-6 observation were performed to spatially resolve the [C\,\textsc{ii}]~158$\,\mu$m line, determine an accurate redshift, and investigate the host galaxy of PSO J083+11 (2019.1.01436.S, PI: I. T. Andika). 
We integrated for a total on-source time of 3145 seconds with observation carried out on October 9, 2019 using ALMA's C43-4 array configuration.
The receivers were set to cover $\sim 258$ GHz, which is the expected [C\,\textsc{ii}] frequency ($\nu_\mathrm{rest} = 1900.5369$~GHz) at $z=6.34$ (see Section \ref{sec:gnirs_spec}). 

The data-set was calibrated with a pipeline implemented in Common Astronomy Software Application\footnote{\url{https://casa.nrao.edu/}} (CASA; \citealt{2007ASPC..376..127M}). 
Next, the \texttt{TCLEAN} task was applied using natural weighting to image the visibilities while maximizing the point source sensitivity. We focused on a 5\arcsec~circular region around the central quasar and performed the cleaning down to 2-sigma. 
Then, we masked out the line-containing channels and used a simple median approximation to model the continuum. After that, we produced continuum-free data by subtracting the continuum model from visibilities.
The final imaged cubes have 30 MHz channel width, a synthesized beam of around $0\farcs42\times0\farcs37$, and a root-mean-square (rms) noise level of $\sim 0.24$ mJy beam$^{-1}$.

\subsection{[C II] spectral profile} 
\label{subsec: cii_spec}

The [C\,\textsc{ii}] spectrum is extracted with an aperture radius of 1\farcs5 (equivalent to 8.3 kpc) to maximize the recoverable emission. We chose this value, because there is no further [C\,\textsc{ii}] flux recovered outside this radius. Ill-defined units issue makes the flux measurement in interferometric maps challenging 
(see \citealt{2019ApJ...881...63N} for further details). 
Assume that we have the aperture fluxes measured inside the dirty image $D$, the clean component $C$ only, and residual image $R$. Then, the scaling factor for correction is $\epsilon = C/(D-R)$. 
The most important factors that governs the scaling factor $\epsilon$ are clean and dirty beam size. In our case, $\epsilon \approx 0.6$.
Then, we use the dirty image flux and multiply it with $\epsilon$ to get the final flux density in proper unit.
Note that if using the final image only to measure the flux, we would obtain $\sim10\%$ larger values. This happens because the final image is a superposition of clean Gaussian components plus residuals.
\begin{figure}[htb!]
    \centering
    \epsscale{1.1}
    \plotone{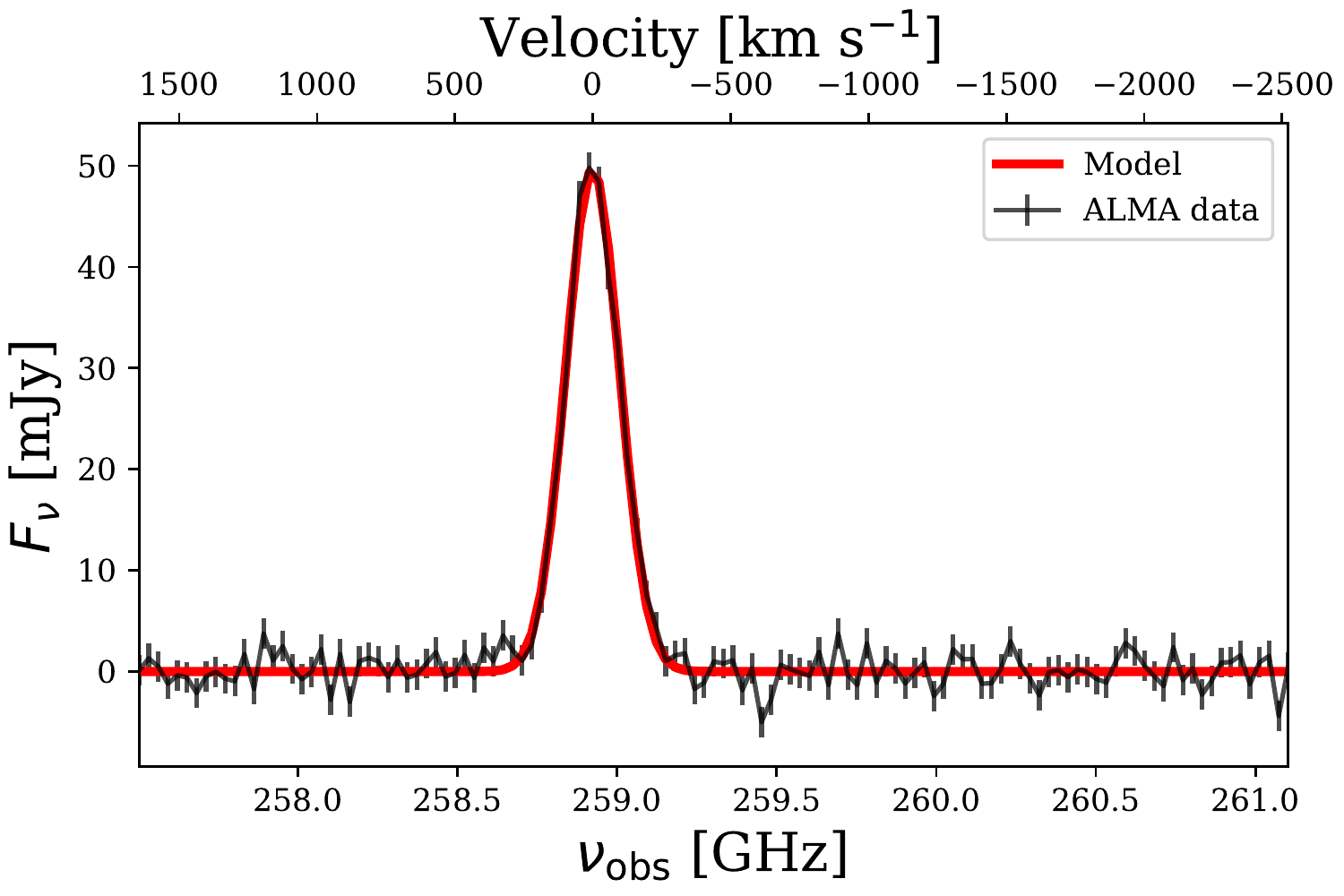}
    \caption{PSO J083+11 continuum-subtracted [C\,\textsc{ii}] spectrum, extracted with 1\farcs5 aperture radius.
    The data (black) is fitted to a Gaussian function (red). The upper axis shows velocities centered at $z = 6.3401$.}
    \label{fig:cii_linefit}
\end{figure}
The extracted spectrum is shown in Figure \ref{fig:cii_linefit} and a Gaussian fit results in the integrated [C\,\textsc{ii}] line flux of of $10.22 \pm 0.35$ Jy km~s$^{-1}$, FWHM of $229 \pm 5$~km~s$^{-1}$, and a redshift of $6.3401 \pm 0.0004$. With respect to [C\,\textsc{ii}] (host galaxy tracer), the Mg\,\textsc{ii} (quasar broad line region tracer) is redshifted by $237 \pm 150$~km~s$^{-1}$. As a comparison, Schindler et al. (in preparation) found that Mg\,\textsc{ii} in high-$z$ quasars can be significantly shifted with respect to the [C\,\textsc{ii}] with a median velocity shift of $-416^{+304}_{-398}$~km~s$^{-1}$.
The Mg\,\textsc{ii} line, which arises from the BLR, may experience strong internal motions or winds, potentially displacing the emission line centers from the systemic redshift \citep{2017ApJ...849...91M}.

\subsection{Moment maps for [C II] and dust continuum emission} 
\label{subsec:moments}

\begin{figure*}[htb!]
    \centering
    \gridline{\fig{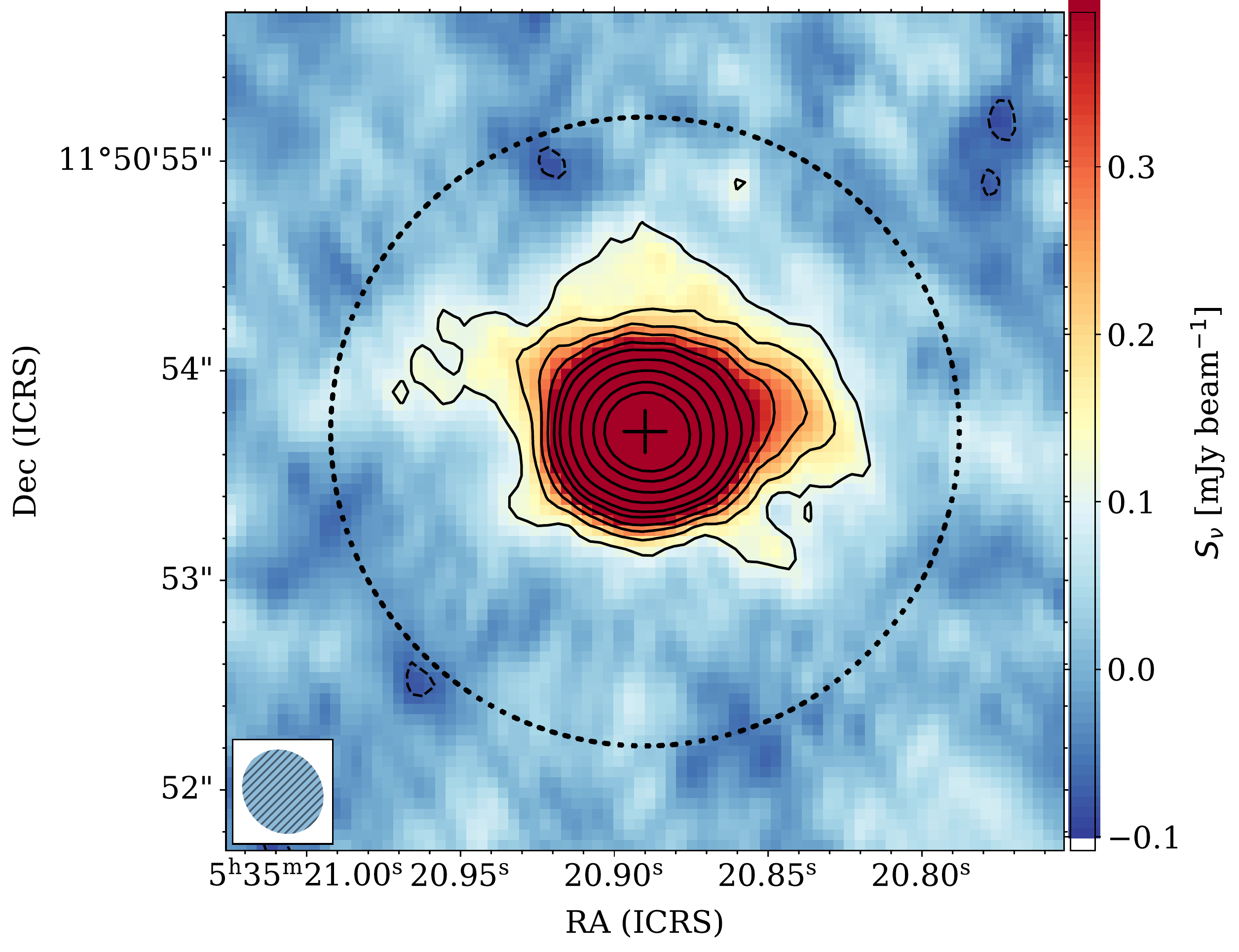}{0.49\textwidth}{(a)}
              \fig{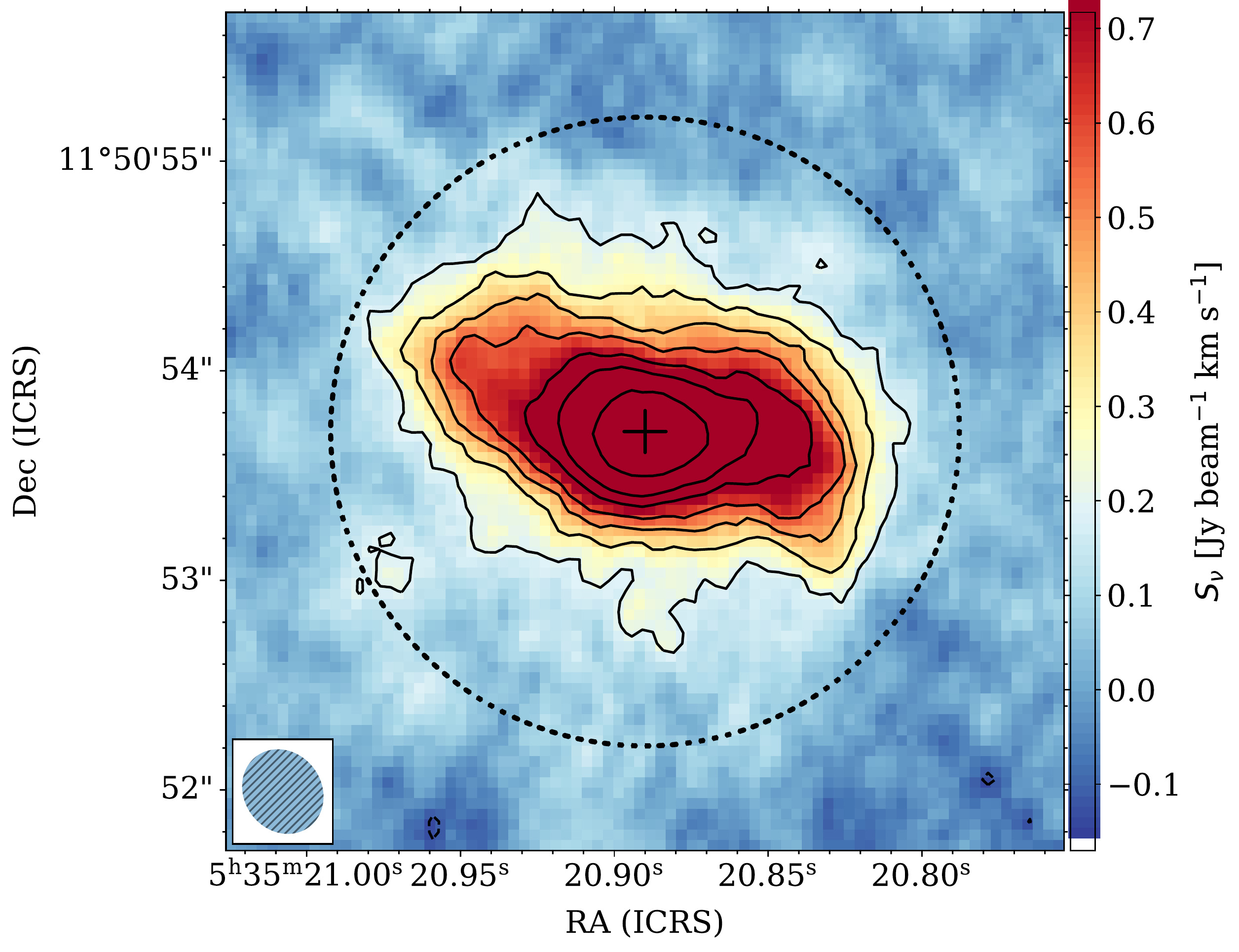}{0.49\textwidth}{(b)}
              }
    \caption{The PSO J083+11's ALMA dust continuum (left) and [C\,\textsc{ii}] velocity-integrated (right) maps. 
    The size of the synthesized beam can be seen at the bottom left of each panel.
    The solid lines in the left panel represent the $[3, 5, 7, 9, 12, 15, 21, 30, 42, 54]\times\sigma$ contours with $\sigma = 0.04$~mJy~beam$^{-1}$ for continuum flux density while right panel shows $\sigma = 0.06$~Jy~beam$^{-1}$~km~s$^{-1}$ for the [C\,\textsc{ii}] velocity-integrated flux.
    Negative contours are shown as dashed lines.
    The  1\farcs5 (8.3 kpc) aperture size that was applied to calculate total flux density is represented by the dotted circles.
    The position of the quasar is marked with a black cross.
    }
    \label{fig:moment_0}
\end{figure*}
We covered two dust continuum spectral windows in our observation, one centered at 244 GHz and the other at 258 GHz.
Pure continuum data are produced by selecting channels which are free from line emission, where we chose effective bandwidth for each spectral window to be around $2500$ km s$^{-1}$. 
Then, we collapsed each spectral window to create moment zero map of the dust continuum.
The final 244 GHz and 258 GHz dust continuum maps have rms noise levels of of 30.7 and 35.8~$\mu$Jy\,beam$^{-1}$, respectively.
Figure \ref{fig:moment_0} shows the continuum map centered at 258 GHz. By using the same circular aperture size as for [C\,\textsc{ii}] -- i.e.\ radius of 1\farcs5 -- we obtain flux densities of $S_{\rm 244\,GHz}=5.10\pm0.15$~mJy and $S_{\rm 258\,GHz}=5.54\pm0.16$~mJy.
The moment zero map of [C\,\textsc{ii}] is also created by collapsing the 700~km~s$^{-1}$ cube width as shown in Figure \ref{fig:moment_0}. This width is equivalent to $3 \times$ [C\,\textsc{ii}] FWHM.
\begin{figure*}[htb!]
    \centering
    \gridline{\fig{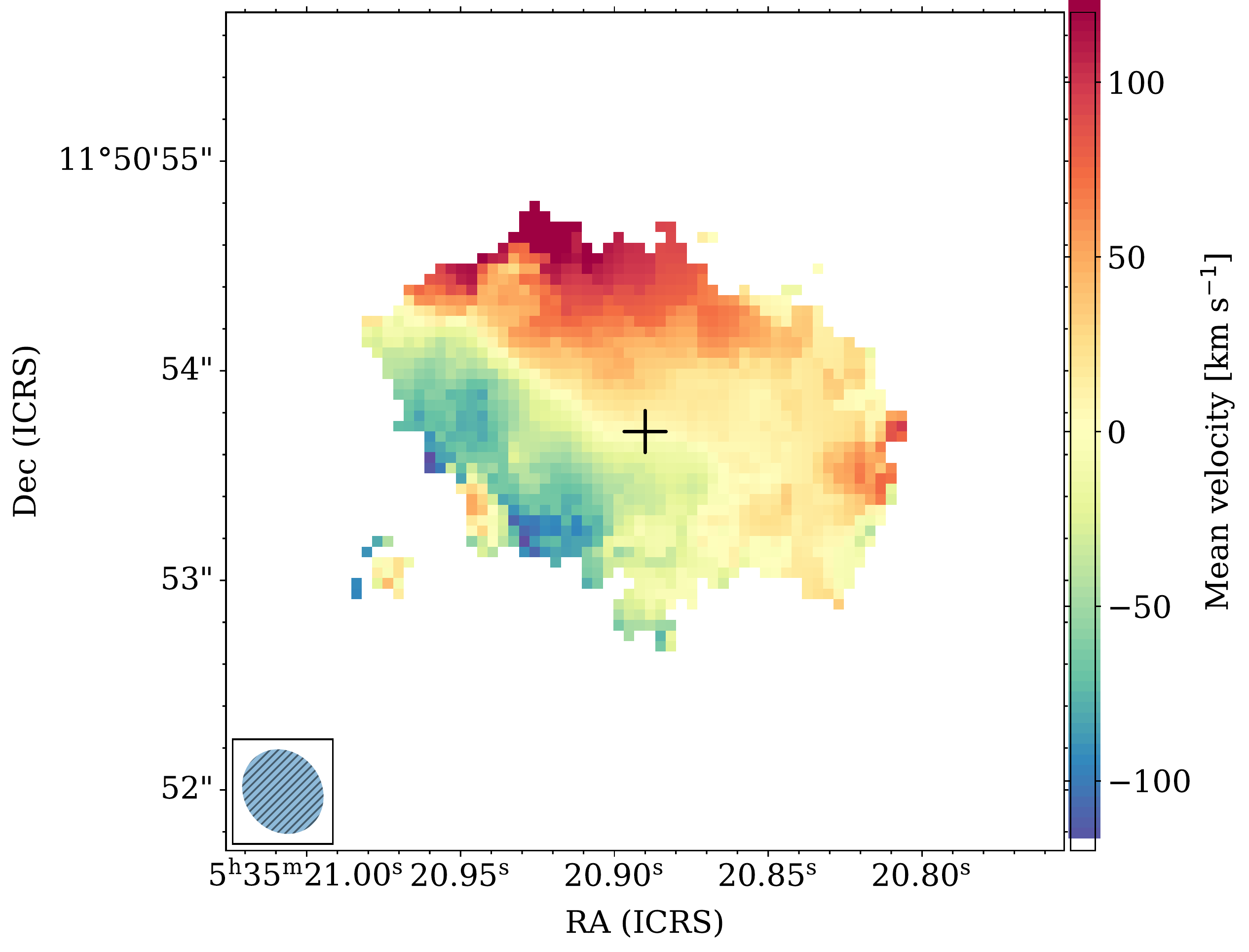}{0.49\textwidth}{(a)}
              \fig{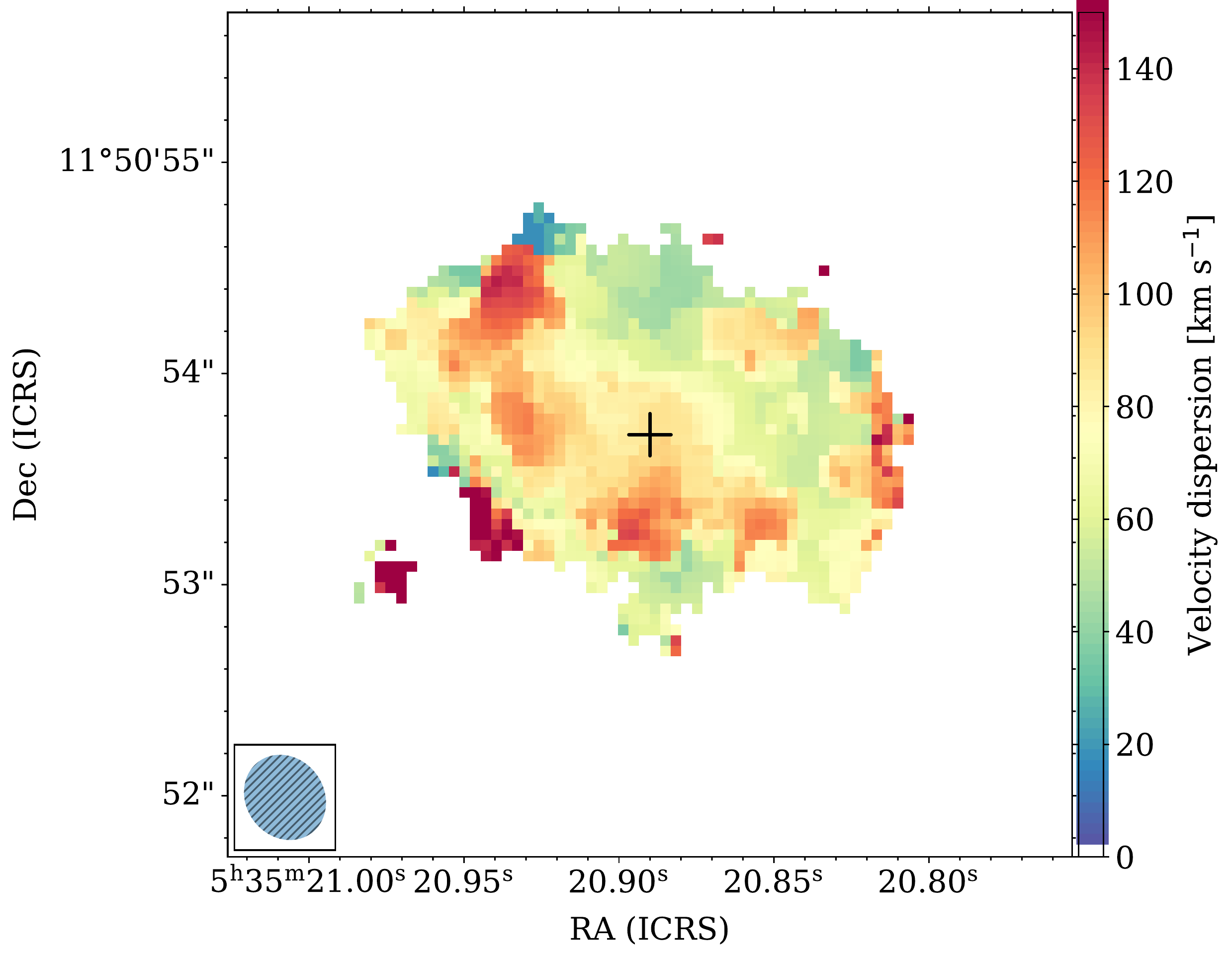}{0.48\textwidth}{(b)}
              }
    \caption{
        Maps of [C\,\textsc{ii}] intensity-weighted velocity (mean velocity field, left panel) and intensity-weighted velocity dispersion (right panel) for PSO J083+11.
        Note that we exclude those pixels with $<3\sigma$ detection in the in the integrated [C\,\textsc{ii}] flux (see Figure \ref{fig:moment_0}). 
        The beam size is shown in the bottom left.
        The position of the quasar is marked with a black cross.
        }
    \label{fig:moment_12}
\end{figure*}

As is visible in Figure \ref{fig:moment_0} both continuum and [C\,\textsc{ii}] emission are spatially resolved with approximate sizes of 2.3~kpc~$\times$~1.7~kpc and 8.2~kpc~$\times$~4.3~kpc, respectively. These sizes are defined as the major and minor axis FWHMs of 2D Gaussian function, which is fitted to those emissions.
In addition, the continuum emission peak overlaps with central [C\,\textsc{ii}] emission.
In Figure \ref{fig:moment_12}, we show the velocity field and dispersion maps of the continuum subtracted [C\,\textsc{ii}] emission\footnote{See the moment map equations and sigma linewidth map definition in \url{https://spectral-cube.readthedocs.io/en/latest/moments.html}}. A mild velocity gradient is seen with the northern part redshifted and the southern part blueshifted. The  morphology in Figure~\ref{fig:moment_0} is seen as single resolved blob while the integrated spectrum (Figure~\ref{fig:cii_linefit}) is represented well by a simple Gaussian profile.

\begin{figure*}[htb!]
    \centering
    \gridline{\fig{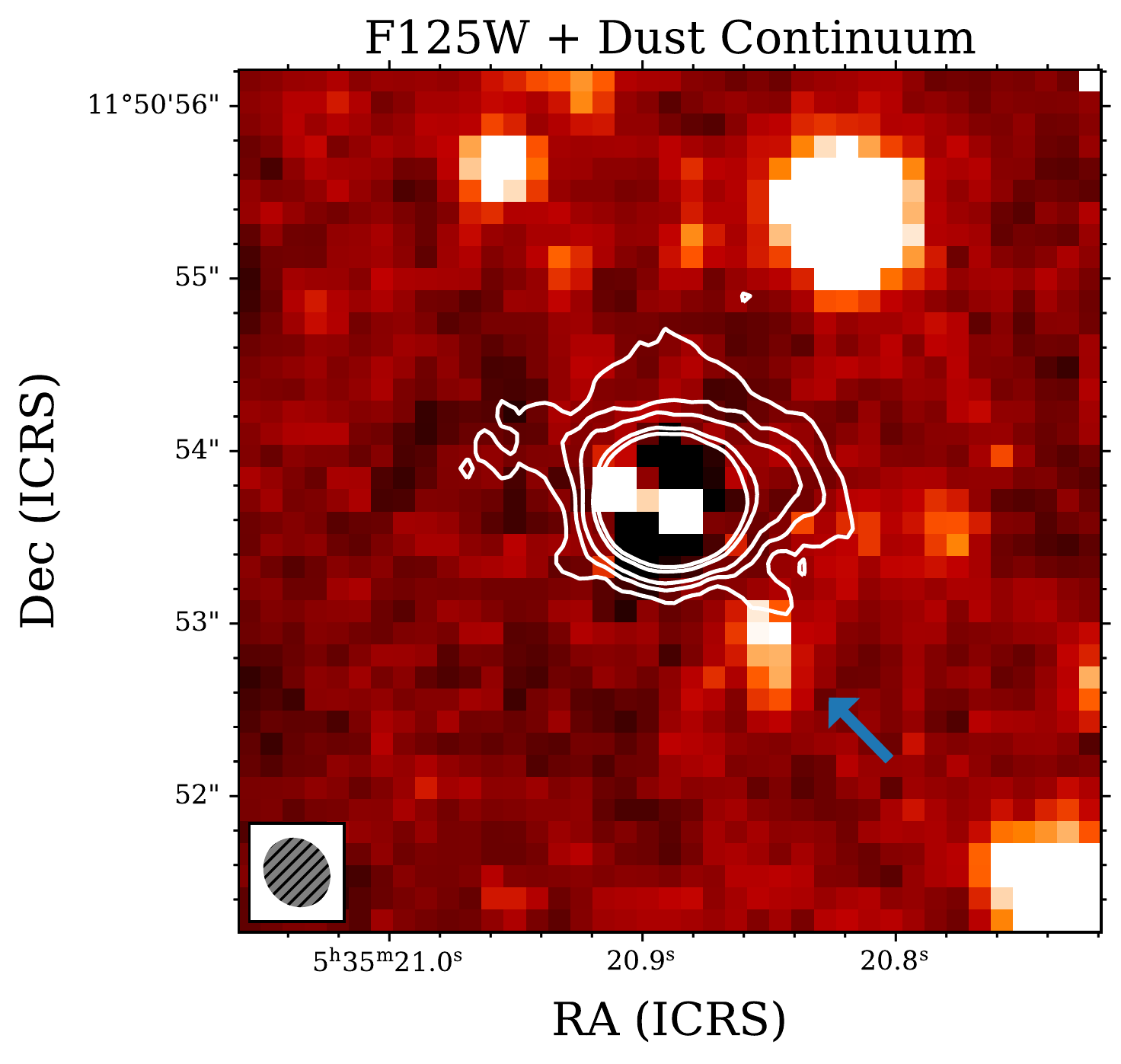}{0.49\textwidth}{(a)}
              \fig{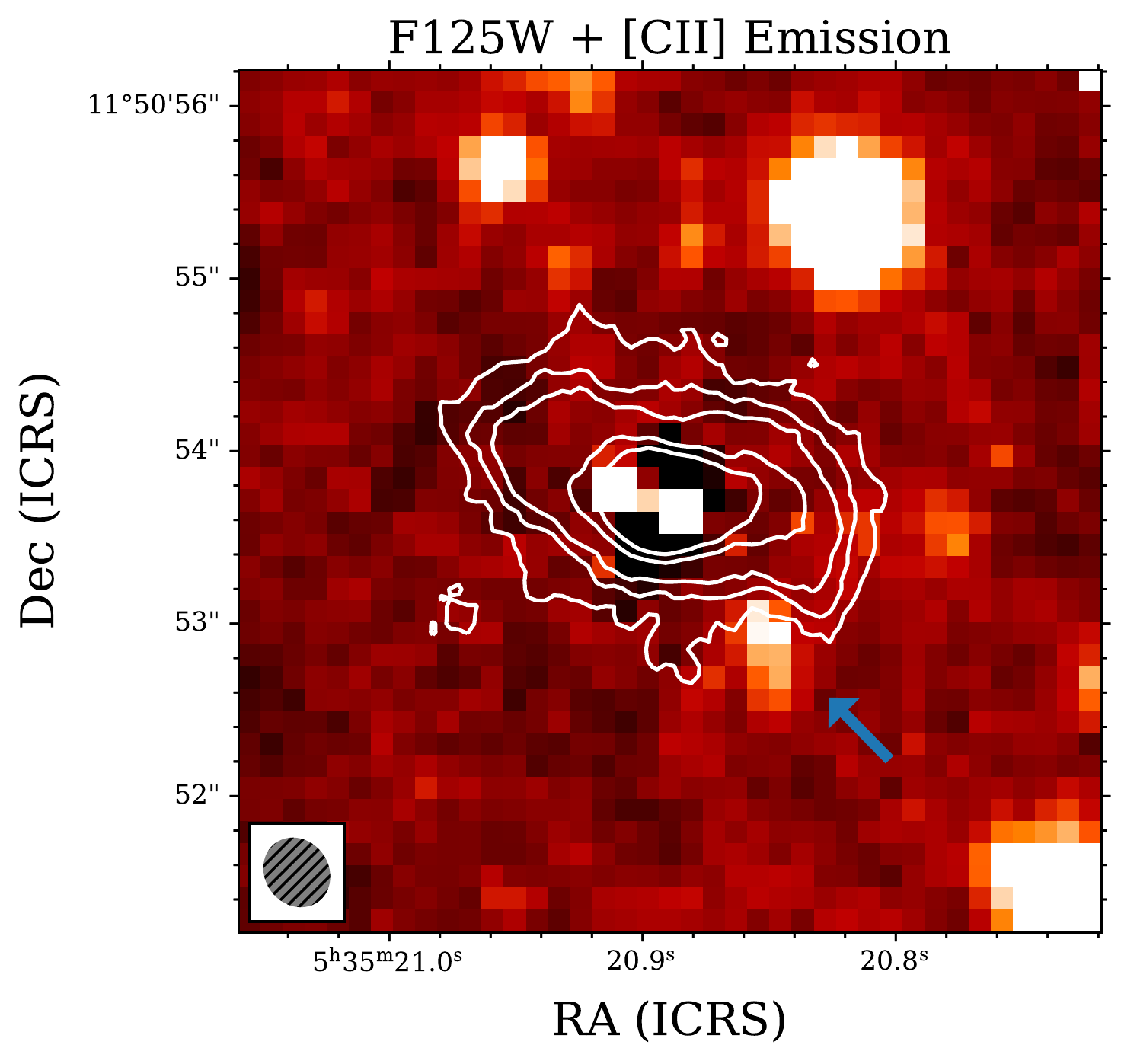}{0.49\textwidth}{(b)}
              }
    \caption{
        PSF-subtracted HST image of PSO J083+11 with an overlay of ALMA dust continuum and [C\,\textsc{ii}] emission maps as contours (from Figure~\ref{fig:moment_0}).
        The white solid lines in the left panel represent the $[3, 5, 7, 12, 15]\times\sigma$ contours with $\sigma = 0.04$~mJy~beam$^{-1}$ for continuum while right panel shows $\sigma = 0.06$~Jy~beam$^{-1}$~km~s$^{-1}$ for the [C\,\textsc{ii}] emission maps.
        Unlike Figure \ref{fig:psf_subtraction}, this PSF-subtracted HST image has been rotated so the North is up and East is left.
        \edit1{The beam size for ALMA data is shown in the bottom left.}
        The neighboring potential  foreground galaxy is located to the south-west from central quasar (blue arrow). There is no apparent dust or [C\,\textsc{ii}] emission at the location of this neighbor, which makes it unlikely to be a physical companion.
        }
    \label{fig:alma_hst}
\end{figure*}

Finally, we overlaid the ALMA on the PSF-subtracted HST image as seen in Figure \ref{fig:alma_hst}.
\edit1{Previously, we identified a potential lensing galaxy as shown in Figure~\ref{fig:psf_subtraction}. However, there is no apparent dust or [C\,\textsc{ii}] emission at the position of this potential lensing companion (see Figure \ref{fig:alma_hst}), which makes it unlikely to be physically associated.}

\subsection{Star formation rate} \label{subsec:sfr_calc}

To derive a star formation rate in the quasar host galaxy we first calculate the [C\,\textsc{ii}]~158\,$\mu$m line luminosity following \cite{2013ARA&A..51..105C}:
\begin{equation}
    \frac{L_{\rm line}}{L_\odot} = 1.04 \times 10^{-3} \frac{S_{\rm line}\Delta v}{\mathrm{Jy~km~s^{-1}}} \left(\frac{D_L}{\mathrm{Mpc}}\right)^2 \frac{\nu_{\rm obs}}{\rm GHz}
\end{equation}
We obtain a luminosity of $L_{\rm [C\,\textsc{ii}]} = (1.04 \pm 0.04) \times 10^{10}\,L_\odot$, which interestingly makes this one of the most luminous [C\,\textsc{ii}] line $z>6$ quasars detected to date \citep{2018ApJ...854...97D,2018ApJ...866..159V}.
The star formation rate (SFR) can be estimated by applying \cite{2014A&A...568A..62D} SFR-$L_{\rm [C\,\textsc{ii}]}$ scaling relations for $z>5$ galaxies:
\begin{equation}
    \frac{\mathrm{SFR}_{\rm [C\,\textsc{ii}]}}{M_\odot \mathrm{yr^{-1}}} = 3.0 \times 10^{-9} \bigg(\frac{L_{\rm [C\,\textsc{ii}]}}{L_\odot}\bigg)^{1.18}
\end{equation}
This will give us value of SFR$_{\rm [C\,\textsc{ii}]} \sim 1990\,M_\odot\,\mathrm{yr^{-1}}$. 
However, one needs to take into account a factor of $\sim 2.5$ for the systematic uncertainty from the scaling relation which makes the possible range of derived SFR for PSO J083+11 SFR$_{\rm [C\,\textsc{ii}]} = 800-4900\,M_\odot\,\mathrm{yr^{-1}}$.

Another important parameter that we can estimate from the ALMA data are the total infrared luminosity ($L_{\rm TIR}$, rest-frame $3-1100$~$\mu$m), far-infrared luminosity ($L_{\rm FIR}$, rest-frame $42.5-122.5$~$\mu$m), and the dust mass, from which we can again calculate an independent star formation rate (e.g.\ \citealt{1988ApJS...68..151H,2012ARA&A..50..531K,2013ARA&A..51..105C}). This can be done by assuming low dust optical depth in the  Rayleigh--Jeans regime, so modeling the SED with a modified blackbody will be sufficient (e.g.\ \citealt{2006ApJ...642..694B,2019ApJ...881...63N}).
The expression is:
\begin{equation}
    S_{\nu_\mathrm{obs}} = f_\mathrm{CMB} (1+z) (D_L)^{-2} \kappa_{\nu_\mathrm{rest}} M_\mathrm{dust} B_{\nu_\mathrm{rest}}(T_{\mathrm{dust},z})
    \label{eq:bb_model}
\end{equation}
where $B_{\nu_\mathrm{rest}}$ is the blackbody radiation function, $D_L$ is luminosity distance, $M_\mathrm{dust}$ is dust mass, while observed and rest frequencies are related as $\nu_\mathrm{rest} = (1+z) \nu_\mathrm{obs}$, and all values are in SI units.
The adopted opacity coefficient following \cite{2003MNRAS.341..589D} and \cite{2019ApJ...881...63N} is:
\begin{equation}
    \kappa_{\nu_\mathrm{rest}} = \kappa_{\nu_0}\left(\frac{\nu_\mathrm{rest}}{\nu_\mathrm{0}}\right)^\beta = 2.64 \left(\frac{\nu_\mathrm{rest}}{c/125\,\mu\mathrm{m}}\right)^\beta ~\mathrm{m^2~kg^{-1}}
\end{equation}
where $c$ is speed of light and $\beta$ is the dust spectral emissivity index. Note that the estimated dust mass will have at least a factor of two for the systematic uncertainty due to the adopted opacity coefficient scaling relation.

Dust heating by the cosmic microwave background (CMB) plays an important role at $z\gtrsim6$ and needs to be taken into consideration following \citep{2013ApJ...766...13D}, namely
\begin{equation}
    f_\mathrm{CMB} = 1 - \frac{B_{\nu_\mathrm{rest}}(T_{\mathrm{CMB},z})}{B_{\nu_\mathrm{rest}}(T_{\mathrm{dust},z})}
\end{equation}
\begin{equation}
    T_{\mathrm{dust},z} = \left(T_\mathrm{dust}^{\beta+4} + T_\mathrm{CMB,z=0}^{\beta+4} \left[(1+z)^{\beta+4} - 1 \right]\right)^{\frac{1}{\beta+4}}
\end{equation}
where $T_\mathrm{dust}$ is the intrinsic dust temperature of the source assuming it is located at redshift zero and $T_\mathrm{CMB} = 2.73(1+z)$\,K is the temperature of the CMB at a given $z$. In our case, $T_\mathrm{CMB} = 20.04$\,K.

Note that we only have 2 data points of continuum measurements -- at 244 and 258 GHz -- both of them located on the Rayleigh--Jeans tail and not reaching the dust SED peak.
This prohibits us to constrain the full SED shape and dust temperature due to degenerated fitting parameters. Hence, further assume that $T_{\mathrm{dust}} = 47$\,K and $\beta=1.6$, which is usually applied for quasar host galaxies at $z\gtrsim6$ (e.g.\ \citealt{2006ApJ...642..694B,2018ApJ...866..159V,2018ApJ...854...97D}).
Scaling the SED model (Equation \ref{eq:bb_model}) to the observed FIR photometry at 244 and 258 GHz results in $M_{\rm dust} = (4.88 \pm 0.14) \times 10^8\,M_\odot$. By integrating the SED, we obtain $L_{\rm FIR} = (1.22 \pm 0.07) \times 10^{13}\,L_\odot$ and $L_{\rm TIR} = (1.72 \pm 0.09) \times 10^{13}\,L_\odot$. This high luminosity means that PSO J083+11 can be classified as an hyper-luminous infrared galaxy (HyLIRG, $L_{\rm TIR}>10^{13}\,L_\odot$).

The TIR luminosity can be converted to a SFR by utilizing the relation from \cite{2011ApJ...737...67M} and \cite{2012ARA&A..50..531K}:
\begin{equation}
    \mathrm{SFR_{TIR}}[M_\odot\,\mathrm{yr^{-1}}] = 1.49 \times 10^{-10} L_{\rm TIR}[L_\odot]
\end{equation}
This gives us $\mathrm{SFR_{TIR}} \sim 2560~M_\odot\,\mathrm{yr^{-1}}$.
Accounting for the factor 3 of systematic uncertainty in the scaling relation, we obtain 1-sigma possible range of $\mathrm{SFR_{TIR}} = 900-7600~M_\odot\,\mathrm{yr^{-1}}$. This is consistent with SFR estimate based on the [C\,\textsc{ii}] luminosity above. A summary of the calculated parameters is shown in Table \ref{tab:phys_par}.

\begin{figure}[htb!]
    \centering
    \epsscale{1.1}
    \plotone{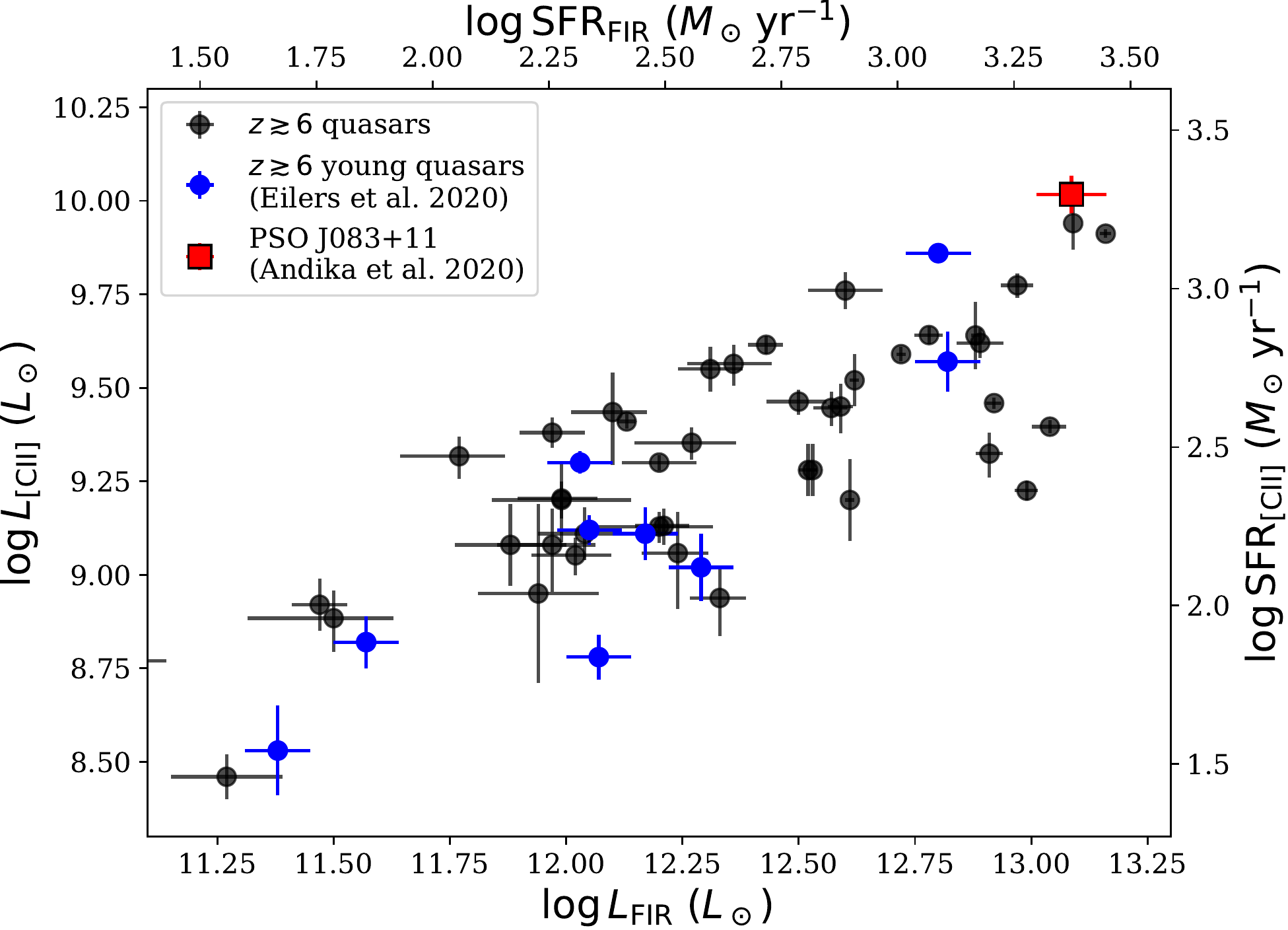}
    \caption{The comparison between the star formation rate estimated from the dust continuum luminosity (following \citealt{2011ApJ...737...67M} and \citealt{2012ARA&A..50..531K}) and the [C\,\textsc{ii}] luminosity (following \citealt{2014A&A...568A..62D}) using assumption of $T_{\mathrm{dust}} = 47$\,K. A sample of $z\gtrsim6$ quasars from literature (black dots, see text) and $z\gtrsim6$ young quasars (blue dots, \citealt{2020arXiv200201811E}) are shown. 
    PSO J083+11 (red square) shows values of $L_{\rm [C\,\textsc{ii}]}$ and $L_{\rm FIR}$ comparable to those high-$z$ quasars with the highest SFR and FIR emission.
    \edit1{For clarity, the error bars of PSO J083+11 have been multiplied by three.}}
    \label{fig:cii_fir}
\end{figure}
To compare PSO J083+11 host galaxies properties with other $z\gtrsim6$ quasars, we took $L_{\rm [C\,\textsc{ii}]}$ and $L_{\rm FIR}$ measurements from literature \citep{2009Natur.457..699W,2012ApJ...751L..25V,2013ApJ...773...44W,2013ApJ...770...13W,2015ApJ...805L...8B,2015ApJ...801..123W,2016ApJ...830...53W,2016ApJ...816...37V,2017ApJ...849...91M,2017ApJ...850..108W,2018ApJ...854...97D} which were recalculated by \cite{2018ApJ...854...97D}. We also added 10 young quasars which was studied by \cite{2020arXiv200201811E}.
As seen in Figure \ref{fig:cii_fir}, PSO J083+11 shows values of $L_{\rm [C\,\textsc{ii}]}$ and $L_{\rm FIR}$ comparable to those high-$z$ quasars with the highest SFR and FIR emission.

\section{Discussion: a Young Weak-Line Quasar?} 
\label{sec:discussion}

Type 1 active galactic nuclei (AGNs) intrinsically exhibit strong broad line emission in the optical and ultraviolet rest-frame regime. However, this is not the case for the weak emission line quasars (WLQs), which show unusually weak or no emission line. It is well established that WLQs are not BL Lacertae objects which usually have spectra dominated by a relatively featureless non-thermal emission continuum due to the effects of the relativistic jet closely aligned to the line of sight of the observer \citep{2009ApJ...699..782D}. Also, the majority of WLQs are radio quiet and they are not related to broad absorption line phenomenon (e.g.\ \citealt{2018MNRAS.479.5075K}). 

\citet{2009ApJ...699..782D} and \citet{2006ApJ...644...86S} established that WLQs are a rare, unique population as their nature could not be explained by gravitational macro- or micro-lensing. In the case of PSO J083+11, our HST observations revealed a potential lensing galaxy located at $\theta\sim1$\,\arcsec\ from the central quasar.
However, the maximum magnification of $\mu \leq 1.07$ does not provide significant boosting to the apparent quasar continuum emission. In addition, we showed that PSO J083+11's UV-to-optical (Figure~\ref{fig:sed_fit}) and far-infrared SED (Section~\ref{subsec:sfr_calc}) is similar to those of ordinary quasars, not a strongly lensed galaxy.
Based on these measurements, the weak-line nature of PSO J083+11 cannot be explained by strong gravitational macro-lensing.
Note that to completely rule out micro-lensing, we would need spectroscopic monitoring and search for the reappearance of the emission lines.
For a stellar lens in a foreground galaxy, the characteristic timescale for micro-lensing is $\sim10$~years \citep{1995ApJ...455...37G}.
Nevertheless, the closest foreground galaxy that we found in HST image is already far enough from the central quasar ($\theta > \theta_\mathrm{E}$, see Section~\ref{sec:hst_result}), making a micro-lensing interpretation is unlikely.

Several theories has been proposed to explain the WLQ phenomenon and according \cite{2015ApJ...805..123P}, they can be classified into two broad categories: (1) soft ionizing continuum idea and (2) anemic broad emission line region model.
In this section we will discuss the most probable explanation of the observed weak-line nature of PSO J083+11.

\subsection{Soft continuum due to super-Eddington accretion?} \label{subsec:no_super_eddington}

One might expect that the BLR is less photoionized so the produced broad emission lines are weak due to soft ionizing continuum. 
How could this happen?
The first proposed mechanism is an intrinsically soft continuum due to a cold accretion disk around a slowly accreting very massive black hole \citep{2011MNRAS.417..681L,2015ApJ...805..123P}, although this possibility is rather low in the case of PSO J083+11. This is because the required critical mass is $M_\mathrm{BH} > 3.6\times10^9~M_\odot$ for a non-rotating ($a=0$) or $M_\mathrm{BH} > 1.4\times10^{10}~M_\odot$ for a nearly maximally rotating ($a=0.998$) black hole while the mass estimated for PSO J083+11 is already a factor $\sim 2$ smaller than the lower of these limits (see Section \ref{sec:bh_properties}). In addition, according to \cite{2013ApJ...775...94V} high-$z$ SMBHs should be rapidly spinning which would yet increase this difference.

The second proposed mechanism is a quasar with a high Eddington ratio is expected to have an optically and geometrically thick inner accretion disk, a so called slim disk \citep{2015ApJ...805..122L}.
The scale height of this thick disk grows as a function of Eddington ratio and becomes a shielding component which prevents the high energy photons from central region to reach and ionize the BLR, leading to observed weak high-ionization line emission \citep{2018MNRAS.480.5184N}. 
This model has been corroborated from observation of WLQs with weak X-ray emission that typically show harder X-ray spectra compared to normal quasars, indicating intrinsic X-ray absorption in these objects \citep{2011ApJ...736...28W,2012ApJ...747...10W}. However, we cannot completely rule out the possibility of the presence of shielding gas between accretion disk and BLR due to the lack of X-ray spectroscopy for PSO J083+11.

The third proposed mechanism is that an extremely high accretion rate could lead to less efficient production of X-rays, which results in an SED that peaks in the ultraviolet \citep{2007ApJS..173....1L,2007ApJ...663..103L}.
This could be related to a quenched X-ray corona, making it smaller in size, or alternatively X-ray photons that are trapped and advected into the black hole before they can diffuse out.
In other words, in these scenarios there would not be enough high-energy photons emitted from the continuum source to produce high-ionization potential lines like C\,\textsc{iv}. However, there would be no problems with producing lower-ionization lines (e.g.\ H$\alpha$, H$\beta$, and Mg\,\textsc{ii}).

The second and third aforementioned mechanisms suggest that on average the low-$z$ WLQs have significantly higher Eddington ratios and luminosities compared to other normal low-$z$ quasars \citep{2014A&A...568A.114M}.
In contrast, we do not see this behavior for PSO J083+11 compared to other high-$z$ quasars because its Eddington ratio is based on the Mg\,\textsc{ii} line and the underlying continuum is not particularly strong ($L_{\rm bol}/L_{\rm Edd} \sim 0.5$; see Figure \ref{fig:masslum}).
However, we have to note that the virial mass calculation relies on the empirical scaling relation derived via reverberation mapping, which might be inadequate in the case of WLQs due to their weakness of emission lines \citep{2015ApJ...805..122L}. The calculated black hole mass tends to be underestimated, making the actual Eddington ratio potentially lower than currently estimated \citep{2019arXiv191006175M}. Moreover,
the typical systematic error of single-epoch virial-mass approach is $\approx 0.4-0.5$~dex \citep{2013BASI...41...61S} and could be larger if the Mg\,\textsc{ii} BLR is complex and may not be virialized yet in this kind of exceptional objects \citep{2015ApJ...805..123P}.

Another potential indicator of extremely high accretion rate is the shape of the continuum itself, as found in a super-Eddington ($L_{\rm bol}/L_{\rm Edd} \geq 9$) quasar, PSO J006+39, at $z\sim 6.6$ with a very blue continuum that was studied by \cite{2019MNRAS.484.2575T}.
This Eddington ratio was estimated by modeling the SED of the UV continuum, where they obtained power-law slope index, $\alpha_\lambda = -2.94 \pm 0.03$. 
This is significantly bluer than the slope of $\alpha_\lambda = -2.33$ predicted from the standard thin disk model \citep{1973A&A....24..337S}.
Although PSO J006+39 is accreting at super-Eddington rate, it is not a WLQ (EW$(\rm C\,\textsc{iv})_{rest} \sim 84$~\AA).
Compared to that object, our measured power-law slope index is consistent with a typical quasar ($\alpha_\lambda = -1.66 \pm 0.01$) and not steeper than those aforementioned models, indicating the absence of super-Eddington accretion.
Hence, overall soft continuum models seem inadequate to explain the weak-line nature of PSO J083+11.

\subsection{Are WLQs young quasars?} 
\label{sec:young_quasar}
With the caveats of all the scenarios presented in the Section \ref{subsec:no_super_eddington}, trying to explain the WLQ nature, another potential explanation for the weak-line nature might lie in gas deficient or anemic BLR clouds, due to the quasar being in the beginning of its accretion phase.
In the first scenario, there might be only a small amount of gas and/or covering factor in the BLR of WLQs \citep{2010ApJ...722L.152S, 2012MNRAS.420.2518N}.
If this is true, all broad lines should have small flux and equivalent widths.
However, unlike normal quasars, WLQs have relatively small EW$(\rm C\,\textsc{iv})_{rest}$/EW$(\rm H\beta)_{rest}$ line ratios which is inconsistent with their BLRs simply having low gas content or small covering factors, although it could play a secondary role \citep{2015ApJ...805..123P}.

On the other hand, the second scenario proposed that in the beginning of an AGN phase, BLRs are still bare because the material from the accretion disk has not yet have sufficient time to reach the region where broad lines will later form \citep{2010MNRAS.404.2028H}.
With the assumption that the wind from the disk has velocities of $\approx 100$ km/s, the time needed to form the BLR is around $\sim 10^3$ years \citep{2010MNRAS.404.2028H}. 
Therefore, if WLQs are indeed an evolutionary quasar phase, they should be rare.
An interesting discovery is that the fraction of WLQs among quasars seems to increase with redshift (e.g.\ \citealt{2009ApJ...699..782D,2016ApJS..227...11B,2019ApJ...873...35S}).
In this picture of a still forming BLR during the WLQs phase, higher-ionization species such as C\,\textsc{iv} would be the weakest because they originate in a region higher above the disk that is not fully formed yet. On the  other hand, low-ionization species (e.g.\ H$\beta$, Mg\,\textsc{ii}) that are formed close to the accretion disk may look normal \citep{2015ApJ...805..123P}.

We should emphasize that most of low-$z$ WLQ studies do not have access to Ly$\alpha$ and the quasar lifetime couldn't be determined with the method that we explained in Section \ref{sec:weak_line}. At $z>6$, the proximity zones are sensitive to the lifetime of the quasars since the intergalactic gas has a finite response time to the quasars' radiation.
The accretion lifetime of PSO J083+11 as derived from its proximity zone gives us a range of $t_\mathrm{Q} = 10^{3.4\pm0.7}$ years, consistent to the BLR formation time at the lower end.
Hence, there is a possibility that the BLR in this object is not fully formed yet, leading to the observed weak emission line signature in the spectrum \citep{2010MNRAS.404.2028H}.
However, as pointed out by \cite{2018ApJ...867...30E}, this quasar could have a higher actual age of substantial accretion compared to the one estimated from its proximity zone size. In that case it could have been growing in a highly obscured phase and the UV continuum radiation had only broken out of this obscuring medium $\sim 10^3$ to $10^{4}$ years before (see also \citealt{2005ApJ...630..705H,2017MNRAS.464.3526D,2018MNRAS.477...45M}).
This might be caused by a huge dust and gas supply at high redshifts, funneled into the center of the host galaxy feeding both star formation and SMBH, but hiding the quasar within it. In line with that, even though tailored at lower-$z$ systems, \cite{1988ApJ...325...74S} argued that ultra-luminous infrared galaxies (ULIRGs) could be the initial stage of a quasar when heavy obscuration was present. Only at the end of this incipient dust-enshrouded phase, the quasar would be revealed in the optical as an unobscured source (e.g.\ \citealt{1988ApJ...328L..35S,1996ARA&A..34..749S,2008ApJS..175..356H}).
A model proposed by \cite{2011ApJ...728L..44L} even predicts that the quasars that have just become unobscured should exhibit no broad emission lines because the BLR would form at a later stage when the dusty torus supplies the fuel to the accretion disk.
The luminous FIR properties that we found in Section \ref{subsec:sfr_calc} would suggest that PSO J083+11 is just at the beginning of its unobscured quasar phase while the host galaxy still experience highly active star formation. This picture is well consistent with the young accretion lifetime derived from the proximity zone size measurement.

\subsection{A small caveat regarding the small proximity zone size} 
\label{sec:dla}

Above we attributed the small proximity zone of PSO J083+11 to a limited unobscured accretion lifetime. There is a hypothetical alternative, the truncation of PSO J083+11's proximity zone due to the presence of an absorption systems within $\leq 10000$\,km\,s$^{-1}$ (or $z > 6.3$) in front of the quasar, just around the edge of the proximity zone. Such a system, like a damped Ly$\alpha$ system (DLA) or Lyman limit system (LLS), would block ionizing radiation from the quasar to the IGM due to its optically thick nature at the Lyman limit \citep{2017ApJ...840...24E,2018ApJ...867...30E,2018ApJ...863L..29D,2019ApJ...885...59B,2019ApJ...887..196F}.

In Figure \ref{fig:abs_sys}, we show hypothetical absorption systems at $z=6.295$ (or $z=2.233$, see below) that might be able to truncate the proximity zone. This redshift value is equivalent to a distance of $R \approx 2.43$~pMpc from the central quasar. For comparison, the proximity zone size that we obtained in Section \ref{sec:weak_line} is $R_p = 1.17\pm 0.32$~pMpc.
The redshift of that system was chosen so that the associated N\,\textsc{v}\,$\lambda\lambda1238,1242$ match the position of the two strong absorption lines observed at 9037\,\AA~and 9066\,\AA. 
However, the positions of other lines that should be present in DLAs (e.g.\ Si\,\textsc{ii}\,$\lambda1260$, O\,\textsc{i}\,$\lambda1302$, Si\,\textsc{iv}\,$\lambda1402$, etc.) do not match any potential absorption lines seen in the observed spectrum.
The presence of low-ionization lines is particularly important because they usually indicate the optically thick self-shielding absorption system \citep{2020arXiv200201811E} which can truncate the proximity zone.
Also, by considering that the inferred distance between absorber and quasar is not really close, the proximity zone might be influenced by such a hypothetical absorber, but would unlikely be significant.

In contrast, there is also an alternative possibility that the two strong absorption lines mentioned before are associated with Mg\,\textsc{ii}\,$\lambda\lambda2796,2803$ from a lower redshift ($z=2.233$) absorption system instead. 
However, we have to note that searching for very weak metal absorption features with our current spectrum is difficult due to the low resolution.
A more thorough analysis to put stringent constraint on potentially associated absorption systems will be done with VLT/MUSE data and we report on this in our next paper (I. T. Andika et al., in preparation).
\begin{figure*}[htb!]
    \centering
    \gridline{\fig{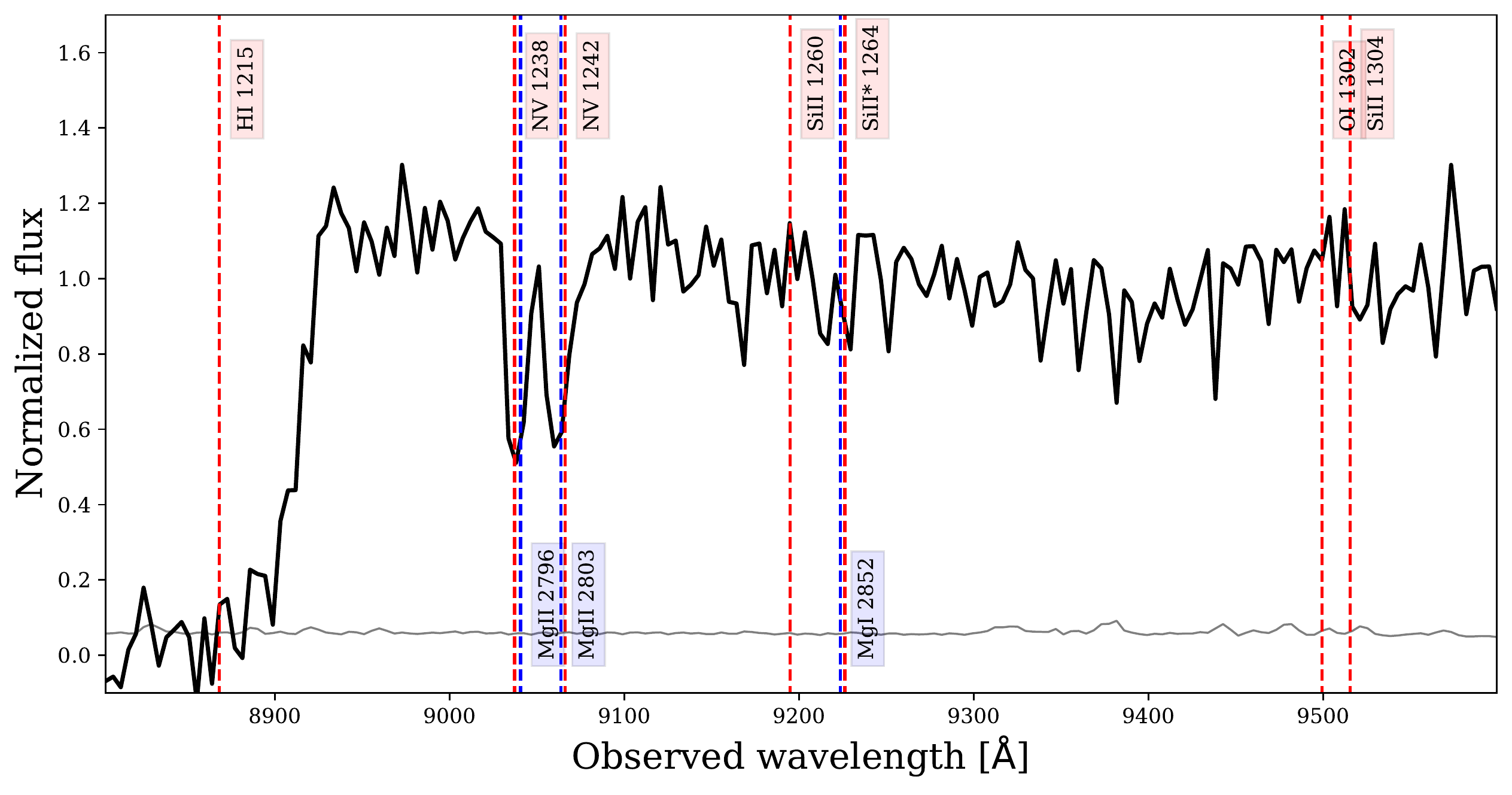}{0.49\textwidth}{(a)}}
    \gridline{\fig{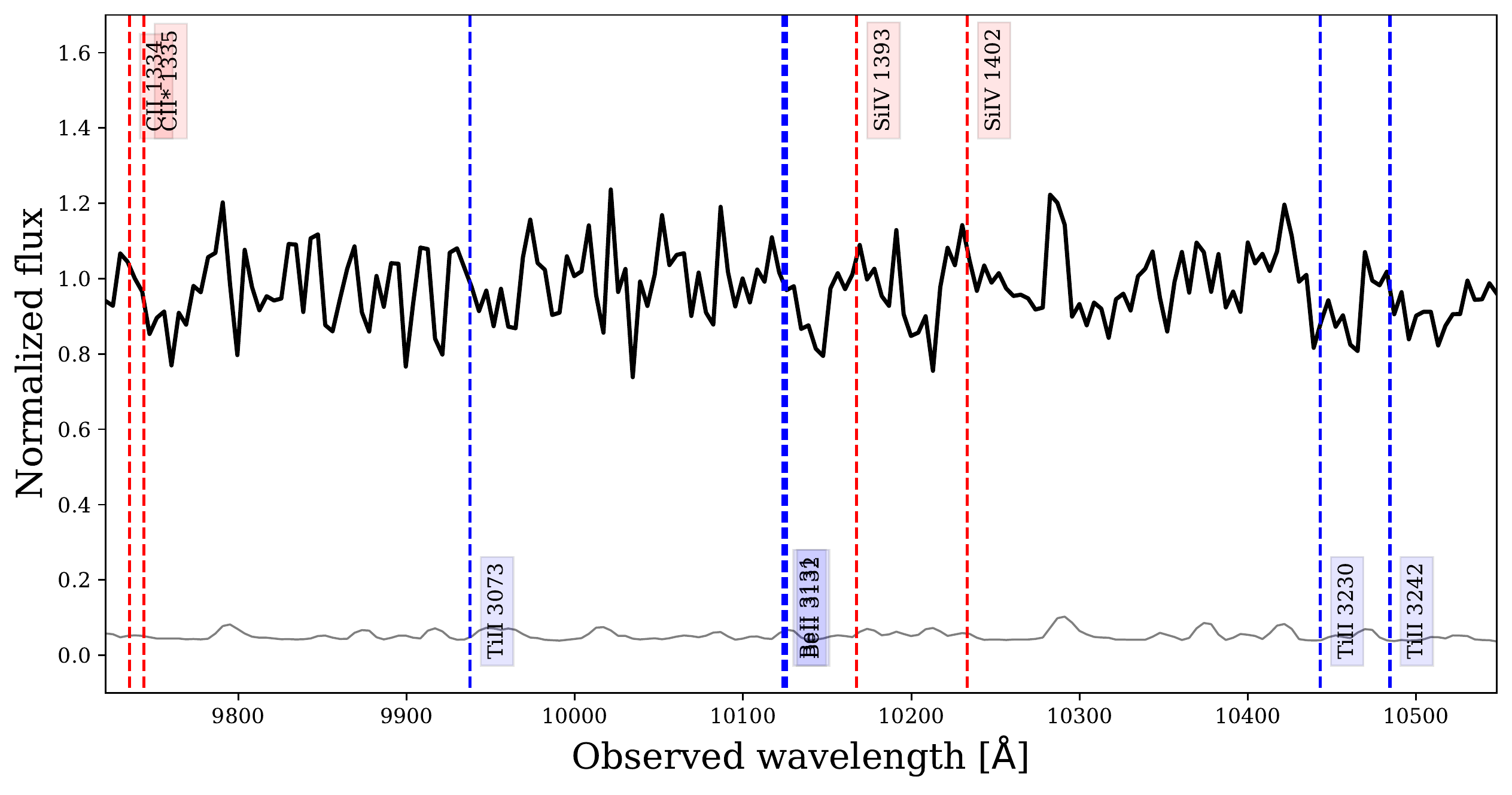}{0.49\textwidth}{(b)}
              \fig{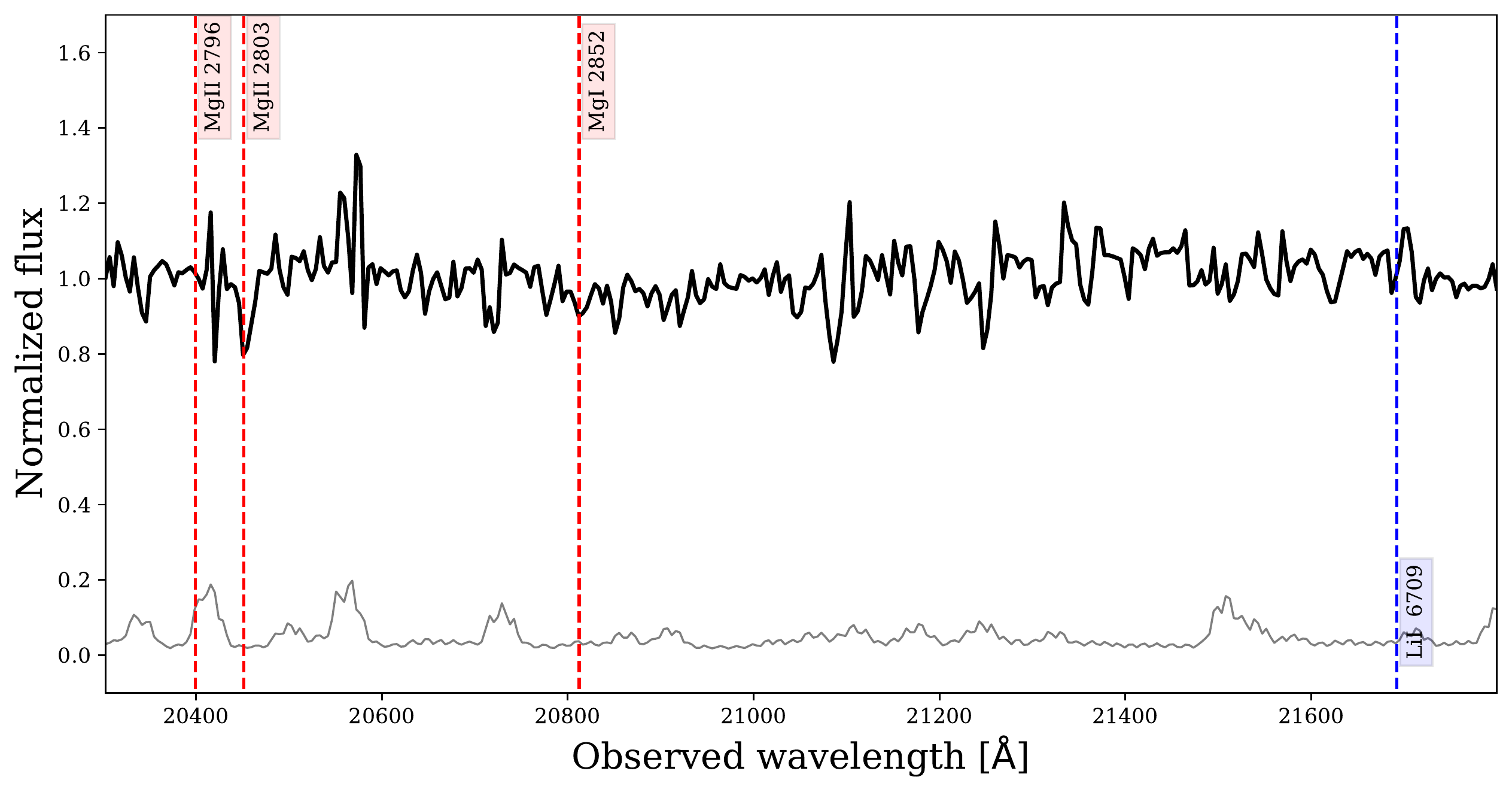}{0.49\textwidth}{(c)}}
    \caption{
    The GNIRS spectrum of PSO J083+11, normalized to the continuum (black). 
    The positions of expected metal lines coming from hypothetical absorption system in front of the quasar at $z=6.295$ and $z=2.233$ are shown in red and blue, respectively. The gray line indicates the noise spectrum. No strong metal lines are found at the expected positions (albeit with low spectral resolution data), discouraging the presence of a DLA or LLS very close to the quasar.
    }
    \label{fig:abs_sys}
\end{figure*}

\section{Summary and Conclusion} 
\label{sec:summary}

In this paper, we presented our effort to increase the number of known quasars at $z>6$, which led to the discovery of PSO J083+11, a new weak emission line quasar at $z=6.3401$. This object was identified using imaging data from PS1, UHS, and unWISE.
This discovery show that our SED-fitting-based quasar selection to identify quasar candidates from large sky imaging surveys is successfully finding quasars, and can be adapted and extended to find larger samples at the highest redshifts.

Our main results on PSO J083+11 can be summarized as follows:

\begin{enumerate}
    \item By using the near infrared spectra from Gemini/GNIRS, we modeled the Mg\,\textsc{ii} emission and the underlying continuum to derive a black hole mass of $\log(M_{\rm BH}) = 9.30^{+0.16}_{-0.10}$~$M_\odot$ and Eddington ratio of $L_{\rm bol}/L_{\rm Edd} = 0.51^{+0.13}_{-0.17}$. This confirms that this object is powered by an actively accreting SMBH with an accretion rate similar to those of most luminous low- and high-$z$ quasars population.
    
    \item Ly$\alpha$+N\textsc{v} emission in this quasar is weak (EW$(\rm Ly\alpha+N\,\textsc{v})_{rest} = 5.65^{+0.72}_{-0.66}$~\AA), suggesting that this is a weak-line quasar. The weak-line nature of Ly$\alpha$+N\,\textsc{v} is not likely caused by IGM strong absorption in the line of sight. This is supported by the absence of C\,\textsc{iv} emission (EW$(\rm C\,\textsc{iv})_{rest} \leq 5.83$~\AA), which suggests the strength of the BLR emission is intrinsically weak.
    The spectrum shows a very small proximity zone $R_{\mathrm p}=1.17\pm0.32$ pMpc which suggests a current quasar lifetime of only $10^3 - 10^{4.5}$ years, at odds with the SMBH mass having formed with the observed instantaneous accretion rate. 

    \item From HST/WFC3 imaging in the F125W filter, we found a potential intervening galaxy located at 1\arcsec~to the southwest direction from the central quasar with a $\mathrm{magnitude} = 25.42 \pm 0.07$. Assuming that it is a star-forming main sequence galaxy at lower redshift and using a point-mass gravitational lens configuration, we find an upper limit of possible lensing magnification $\mu \leq 1.07$, implying no relevant effect of boosting to quasar apparent emission. The quasar is also observed as a pure point-source with no additional emission component.

    \item ALMA band-6 observations for PSO J083+11 have detected both dust continuum and spatially resolved [C\,\textsc{ii}] emission from the host galaxy. We derived an accurate redshift from [C\,\textsc{ii}] ($z_{\rm [C\,\textsc{ii}]} = 6.3401 \pm 0.0004$). The resolved extended morphology of this line might be caused by a reminiscent of a merger or in any case unrelaxed ongoing formation. This quasar is among the most luminous [C\,\textsc{ii}] emitters to date.
    Modeling the Rayleigh-Jeans tail of dust continuum with a modified blackbody function give us a constrain on dust mass of $M_{\rm dust} = (4.88 \pm 0.14) \times 10^8~M_\odot$ and star formation rate of SFR$_{\rm TIR} = 900-7600~M_\odot\,\mathrm{yr^{-1}}$, similar to that of HyLIRGs. This value is also consistent with the SFR derived from the [C\,\textsc{ii}] emission (SFR$_{\rm [C\,\textsc{ii}]} = 800-4900~M_\odot\,\mathrm{yr^{-1}}$).
    
    \item Considering the quasar lifetime of PSO J083+11 and BLR formation timescale, we propose that the weak emission line profile in our young quasar spectrum is caused by a BLR that is not fully formed yet rather than continuum boosting by gravitational lensing or soft continuum emission due to super-Eddington accretion. However, we cannot completely rule out the possibility of the presence of shielding gas between accretion disk and BLR due to the lack of X-ray spectroscopy. 

\end{enumerate}

Overall, this quasar paints a puzzling picture of a supermassive black hole, at odds with seemingly only a very recent onset of currently moderate accretion.
In the future, a thorough search for potentially  associated absorption systems which could affect the accuracy of our quasar lifetime estimation will be done with VLT/MUSE data. Very beneficial would be the availability of X-ray spectra from facilities like XMM-Newton or Chandra to get better constraints on the SED modeling, accretion rate, and to check the possibility of shielding gas possibly present between accretion disk and BLR.

In context of an early phase of quasar (re)activation, the narrow emission line region will take longer time to form compared to broad emission line region and is likely to be absent in this particular stage \citep{2010MNRAS.404.2028H}. Future facilities like the James Webb Space Telescope (JWST) will enable us to observe $z>6$ WLQs in the mid-infrared.
Hence, we can test for this hypothesis by constraining the NLR properties and tracing the presence of this region by studying the [O\,\textsc{iii}]~$\lambda5007$ line profile.
Further statistical studies, supported by a larger sample of WLQs at highest accessible redshifts are required for establishing connections between normal quasars, young quasars, and those which are weak-lines.
This will play an important role in our understanding of quasar evolution, the rapid formation of the first supermassive black holes, and structure formation in general in the Universe at the dawn of cosmic time.

\acknowledgments

We thank the anonymous referees for constructive comments on the manuscript.
We would like to thank Dian P. Triani for the fruitful discussions on modeling the star-forming main sequence galaxy and Anton T. Jaelani for the valuable insight regarding the gravitational lensing model.
We gratefully acknowledge the assistance of Joseph F. Hennawi and Iskren Georgiev on reducing the spectra and correcting the telluric absorption.
I. T. Andika is extremely grateful to Adriana Larasati and Fenyka A. Jeanuarieke, who give unending encouragement and always being there, especially in the most difficult moments.

F. Walter, B. P. Venemans, M. Novak and M. Neeleman acknowledge the support from the ERC Advanced Grant 740246 (Cosmic Gas).
A.-C. Eilers acknowledges support by NASA through the NASA Hubble Fellowship grant $\#$HF2-51434 awarded by the Space Telescope Science Institute, which is operated by the Association of Universities for Research in Astronomy, Inc., for NASA, under contract NAS5-26555.
I. T. Andika acknowledges support from the International Max Planck Research School for Astronomy and Cosmic Physics at the University of Heidelberg (IMPRS-HD).
This work was supported by the Deutsches Zentrum für Luft- und Raumfahrt (DLR) under grant number 50~OR~2001.

The Pan-STARRS1 Surveys (PS1) and the PS1 public science archive have been made possible through contributions by the Institute for Astronomy, the University of Hawaii, the Pan-STARRS Project Office, the Max-Planck Society and its participating institutes, the Max Planck Institute for Astronomy, Heidelberg and the Max Planck Institute for Extraterrestrial Physics, Garching, The Johns Hopkins University, Durham University, the University of Edinburgh, the Queen's University Belfast, the Harvard-Smithsonian Center for Astrophysics, the Las Cumbres Observatory Global Telescope Network Incorporated, the National Central University of Taiwan, the Space Telescope Science Institute, the National Aeronautics and Space Administration under Grant No. NNX08AR22G issued through the Planetary Science Division of the NASA Science Mission Directorate, the National Science Foundation Grant No. AST-1238877, the University of Maryland, Eotvos Lorand University (ELTE), the Los Alamos National Laboratory, and the Gordon and Betty Moore Foundation.

This paper makes use of the following ALMA data: ADS/JAO.ALMA\#2019.1.01436.S. ALMA is a partnership of ESO (representing its member states), NSF (USA) and NINS (Japan), together with NRC (Canada), MOST and ASIAA (Taiwan), and KASI (Republic of Korea), in cooperation with the Republic of Chile. The Joint ALMA Observatory is operated by ESO, AUI/NRAO and NAOJ.

Part of data presented in this paper is based on observations obtained at the international Gemini Observatory (GN-2019A-FT-204). Gemini Observatory is managed by the Association of Universities for Research in Astronomy (AURA) under a cooperative agreement with the National Science Foundation. on behalf of the Gemini Observatory partnership: the National Science Foundation (United States), National Research Council (Canada), Agencia Nacional de Investigaci\'{o}n y Desarrollo (Chile), Ministerio de Ciencia, Tecnolog\'{i}a e Innovaci\'{o}n (Argentina), Minist\'{e}rio da Ci\^{e}ncia, Tecnologia, Inova\c{c}\~{o}es e Comunica\c{c}\~{o}es (Brazil), and Korea Astronomy and Space Science Institute (Republic of Korea).

This paper includes data gathered with FIRE at 6.5 meter Magellan Baade Telescopes located at Las Campanas Observatory.

We acknowledge the use of the DELS, DES, VHS, and UKIDSS, UHS, and WISE data.


\vspace{5mm}

\facilities{HST (WFC3, ACS), Magellan:Baade (FIRE), Gemini:Gillett (GNIRS), ALMA, NTT (SOFI), PS1 (GPC1), UKIRT, WISE, ESO:VISTA}


\software{Astropy \citep{2013A&A...558A..33A,2018AJ....156..123A},
Lmfit \citep{2019zndo...3381550N},
APLpy \citep{aplpy2012,aplpy2019},
spectral-cube \citep{2016ascl.soft09017R},
SciPy \citep{2020SciPy-NMeth}
}

\vspace{30mm}


\clearpage

\appendix

\section{Spectroscopically Rejected Candidates}
\label{appendix_a}

The spectroscopically rejected candidates that we found in our follow-ups are reported. 
We adopted the International Astronomical Union naming convention for these sources, which is ``PSO JRRR.rrrr+DD.dddd'', where RRR.rrrr and +DD.dddd are the right ascension and declination in decimal degrees (J2000), respectively.
The names, PS1 $z$-band magnitudes ($z_\mathrm{PS1}$), PS1 $y$-band magnitudes ($y_\mathrm{PS1}$), and VHS $J$-band magnitudes ($J_\mathrm{VHS}$) are reported in Table \ref{tab:reject}.
An accurate spectral classification of the sources is beyond the scope of this work.
\begin{deluxetable}{ccccc}[htb!]
	\tablenum{3}
	\tablecaption{Spectroscopic rejected of candidates which are definitely not $z>6$ quasars.}
	\label{tab:reject}
	\tablehead{
		\colhead{Name} & \colhead{$z_\mathrm{PS1}$} & \colhead{$y_\mathrm{PS1}$} & \colhead{$J_\mathrm{VHS}$}
	}	
	\startdata
	PSO J065.5314$-$13.3353 & $22.06\pm0.18$ & $20.54\pm0.11$ & $20.54\pm0.16$ \\
	PSO J134.2027$-$07.1366 & $22.03\pm0.26$ & $20.30\pm0.09$ & $19.73\pm0.10$ \\
	PSO J123.0135$-$01.9006 & $21.93\pm0.19$ & $20.44\pm0.10$ & $19.96\pm0.19$ \\
	PSO J002.1774$-$02.9102 & $22.83\pm0.30$ & $21.06\pm0.13$ & $20.19\pm0.15$ \\
	PSO J303.7815$-$00.4066 & $22.01\pm0.15$ & $20.49\pm0.10$ & $19.97\pm0.13$ \\
	\enddata
\end{deluxetable}

\clearpage

\section{PSO J344.1442--02.7664: A new quasar at redshift \texorpdfstring{$\sim6.5$}{z=6.5}}
\label{appendix_b}

\begin{figure*}[htb!]
    \centering
    \plotone{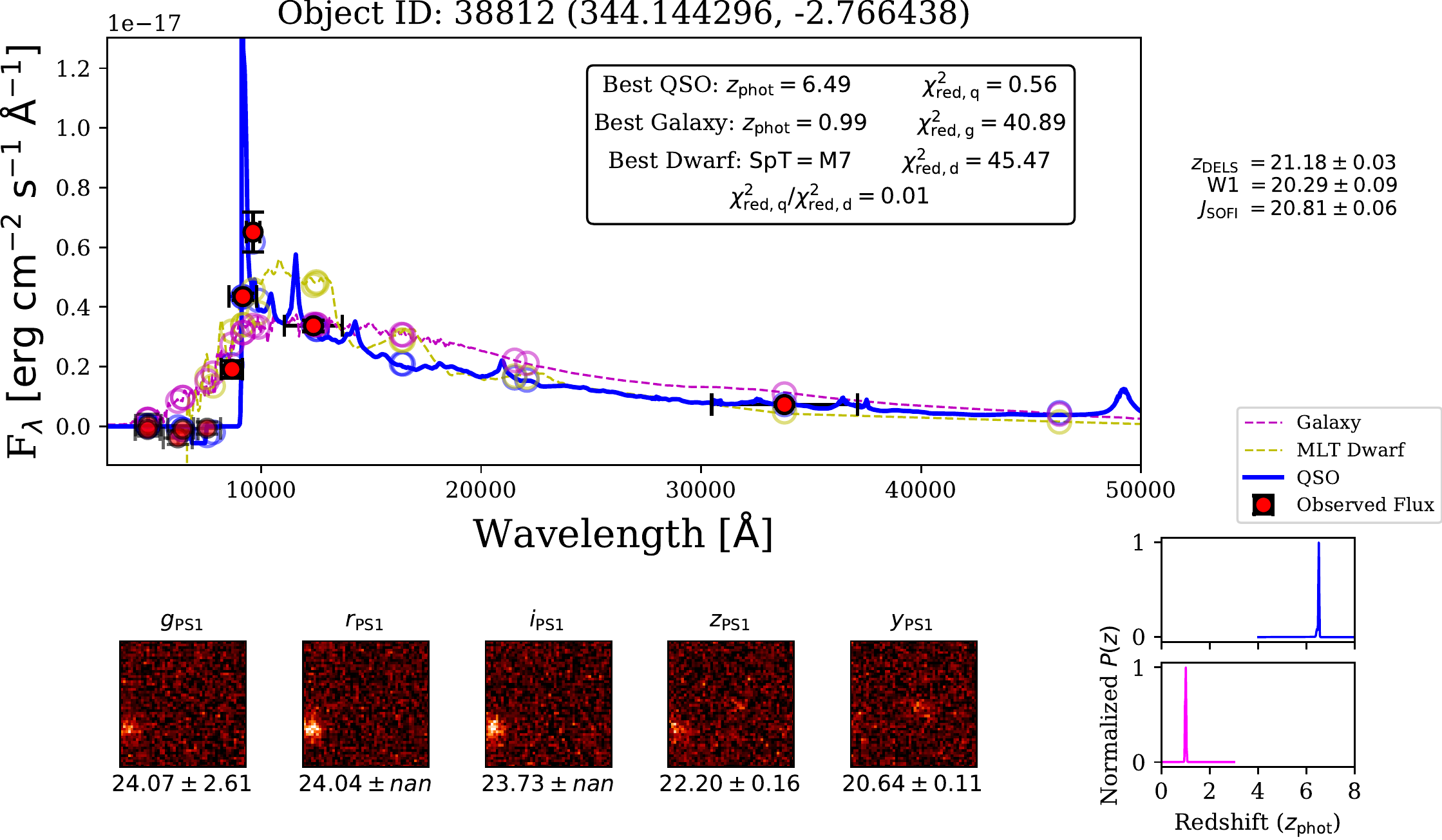}
    \caption{SED fitting result for PSO J344.1442--02.7664. Photometry data is shown with red filled circles with error bars in the top panel. The best-fit quasar spectral template is shown with the blue line and blue circles for model photometry. The same goes for galaxy (magenta) and MLT dwarf models (yellow).
    The bottom panels show 12\arcsec~cutouts in the 5 PS1 bandpasses. All written magnitudes are corrected for Galactic extinction.
    Finally, the bottom right panel shows the probability density function (PDF) of calculated photo-$z$'s for quasar (blue line) and galaxy (magenta line) models.}
    \label{fig:sed_fit_2}
\end{figure*}

\begin{figure*}[htb!]
    \centering
    \plotone{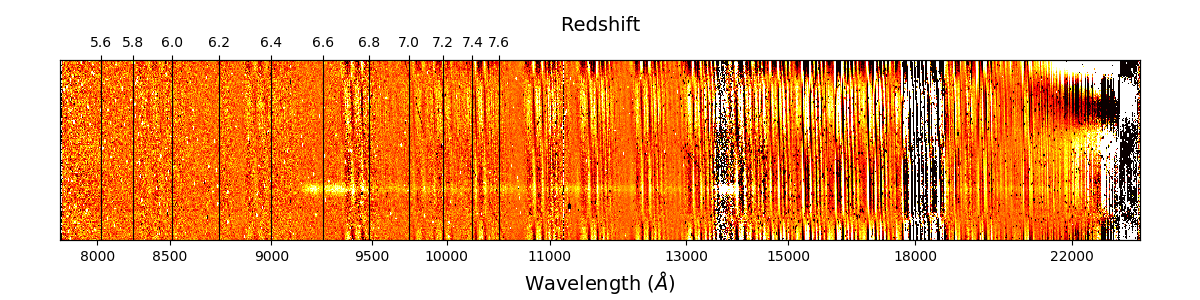}
    \caption{Two-dimensional spectrum of PSO J344.1442--02.7664 obtained with Magellan/FIRE. A strong Ly$\alpha$ break is clearly seen around the observed-frame wavelength of 9100 \AA~which means that the Ly$\alpha$ emission is redshifted to $z\sim6.5$.}
    \label{fig:fire_2d}
\end{figure*}

We report the discovery of PSO J344.1442--02.7664, the second quasar that we found within the PS1, DELS, and unWISE catalogs.
The $J$-band near-infrared (NIR) photometry of this object was obtained by using the NTT/SofI \citep{1998Msngr..91....9M} with exposure time of 15 minutes on July 16, 2019. 
The data were reduced using standard procedures (see \citealt{2014AJ....148...14B} for details).
Then, we did low-resolution NIR spectroscopic follow-up by using 6.5m-Magellan/FIRE on August 9, 2019. 
The quasar was observed for 10 minutes with high-throughput prism mode ($R=500$) using 0\farcs6 slit width. This in principle gives us $R = 500$ spectral resolution with the wavelength coverage of 0.82--2.51\,$\mu$m.
The photometric data and the SED fitting results can be seen in Figure \ref{fig:sed_fit_2}. 
On the other hand, Figure \ref{fig:fire_2d} shows the two-dimensional spectrum where we estimated the redshift of $z\sim6.5$ based on a strong Ly$\alpha$ break around the observed-frame wavelength of 9100 \AA.
The current low-resolution spectrum is not sufficient to calculate accurate emission line and black hole properties for this particular object. Hence, the detailed analysis will be reported later in our next paper (Andika et al., in preparation).

\clearpage

\bibliography{biblio}{}
\bibliographystyle{aasjournal}



\end{document}